# RESEARCH ARTICLE

# Predictability of real temporal networks

Disheng Tang[1,2,3], Wenbo Du[1,3], Louis Shekhtman[4], Yijie Wang[1,3], Shlomo Havlin[5], Xianbin Cao[1,3,*] and Gang Yan[2,6,7,*]

[1]School of Electronic and Information Engineering, Beihang University, Beijing 100191, China; [2]School of Physics Science and Engineering, Tongji University, Shanghai 200092, China; [3]National Engineering Laboratory of Big Data Application Technologies of Comprehensive Transportation, Beijing 100191, China; [4]Network Science Institute, Northeastern University, Boston, MA 02115, USA; [5]Department of Physics, Bar Ilan University, Ramat Gan 5290002, Israel; [6]Shanghai Institute of Intelligence Science and Technology, Tongji University, Shanghai 200092, China and [7]CAS Center for Excellence in Brain Science and Intelligence Technology, Chinese Academy of Sciences, Shanghai 200031, China

*Corresponding authors. E-mails: xbcao@buaa.edu.cn; gyan@tongji.edu.cn

## ABSTRACT

Links in most real networks often change over time. Such temporality of links encodes the ordering and causality of interactions between nodes and has a profound effect on network dynamics and function. Empirical evidence has shown that the temporal nature of links in many real-world networks is not random. Nonetheless, it is challenging to predict temporal link patterns while considering the entanglement between topological and temporal link patterns. Here, we propose an entropy-rate-based framework, based on combined topological–temporal regularities, for quantifying the predictability of any temporal network. We apply our framework on various model networks, demonstrating that it indeed captures the intrinsic topological–temporal regularities whereas previous methods considered only temporal aspects. We also apply our framework on 18 real networks of different types and determine their predictability. Interestingly, we find that, for most real temporal networks, despite the greater complexity of predictability brought by the increase in dimension, the combined topological–temporal predictability is higher than the temporal predictability. Our results demonstrate the necessity for incorporating both temporal and topological aspects of networks in order to improve predictions of dynamical processes.

## INTRODUCTION

Link temporality describes the time-varying nature of couplings and interactions between nodes in real networks [1–12], which has been found to significantly affect network dynamics. Examples include innovative or epidemic diffusion [13], information aggregation [14], the emergence of cooperation [15] and the achievability of control [16]. Hence, in order to alter network dynamical states in a desirable way, it is essential to quantitatively understand both topological and temporal patterns. This raises a fundamental question: how predictable are real temporal networks? This question is much broader and distinct from time-series forecasting [17–19], which aims to predict the future evolution of *single* variables, and link prediction [20–23], the goal of which is to uncover the missing or future links in *static* networks. Here we offer an entropy-rate-based framework that considers the combined topology–temporal patterns and apply them to a wide range of model and real weighted and unweighted temporal networks, uncovering the prediction limits of real temporal networks.

## ANALYTICAL FRAMEWORK FOR PREDICTABILITY

A temporal network with $n$ nodes consists of a series of snapshots (Fig. 1A), which can be described as a 2D expanded matrix $M$ (Fig. 1B). Each column in $M$ represents one snapshot and each row represents the temporality of a possible link, i.e. whether this link is present in a specific snapshot and its weight. Since all pairs of nodes must be taken into account, the number of rows in $M$ is $n^2$. The full information of the temporal network is encoded by this matrix $M$, which can be viewed as a stochastic vector process—a sequence of random vectors. To quantify the predictability of this vector process, we use the entropy rate, $H$, i.e. the asymptotic lower bound on the per-symbol description length [24], which is a rigorous measure of the level of randomness in the process. As illustrated in Fig. 1C, $H$ can be calculated using a generalized Lempel–Ziv algorithm [25], of which the essence is to calculate the recurrence times of different patterns within a square: a 2D square with side $k$ is defined as $M_{C(k)}$, where $C(k) = \{v = (t, s) \in Z^2 : 0 \le t \le k, 0 \le s \le k\}$ and $v$ denotes the coordination of an element in $M$; $\Lambda_v^v$ represents the smallest integer $k$ such that block $M_{v-C(k)}$ does not occur within the rectangle $(\mathbf{0},\ v]$ except at position $v$, where $\mathbf{0} = (0,\ 0)$. It has been proven [25] that $\liminf_{n\to\infty} \frac{n^2 \log n^2}{\sum_{v\in C(n)} (\Lambda_v^v)^2} \to H$. Thus, the entropy rate $H$ of the matrix $M$ is captured by

$$H(M) = \frac{n^2 \log n^2}{\sum_{v\in C(n)} (\Lambda_v^v)^2}, \qquad (1)$$

when the temporal network has a large number of snapshots (see Supplementary Material, Section II).

The predictability of a temporal network is the probability $\Pi_M$ that a predictive algorithm can correctly forecast the future evolution of this network based on its history. Once we have the entropy rate $H(M)$, the upper bound of predictability $\Pi_M^{\max}$ can be obtained by solving

$$\begin{aligned} H(M) = &-\left(\Pi_M^{\max} \log(\Pi_M^{\max}) \right. \\ &\left. + (1-\Pi_M^{\max}) \log(1-\Pi_M^{\max})\right) \\ &+ (1-\Pi_M^{\max}) \log(N-1), \qquad (2) \end{aligned}$$

where $N$ is the number of unique values in matrix $M$ (see the 'Methods' section and Supplementary Material, Section III). Here, $\Pi_M^{\max}$ is the fundamental limit of predictability, i.e. in principle, no algorithm can predict the temporal network with an accuracy higher than $\Pi_M^{\max}$. It is worth stressing that the entropy rate obtained with the generalized Lempel–Ziv algorithm is an asymptotic measure of randomness and Eq. (1) becomes more accurate when the number of time steps is larger. Hence, given the finite number of snapshots in real temporal network data sets, the calculated value of $\Pi_M^{\max}$ should be interpreted as an asymptotic estimate of the upper bound of predictability. Moreover, $\Pi_M^{\max}$ is an intrinsic property of the temporal network and does not depend on a specific predictive algorithm.

The snapshots of real temporal networks are usually very sparse, so most rows in $M$ consist of many zeros. Thus, we sort the rows, i.e. all potential links, in descending order according to the number of their occurrences in all snapshots and remove those links that are present in <10% of the snapshots, obtaining a new matrix $\widetilde{M}$ (see the 'Methods' section). Our analyses in both model and real networks show that the filtering process and the ordering of rows in $\widetilde{M}$ have a negligible effect on the predictability (see Fig. 1D, and also Supplementary Material, Sections IV and V); therefore, we use $\Pi_{\widetilde{M}}^{\max}$ to quantify the predictability of temporal networks hereafter. Note that the original entropy rate (Eq. 1) applies to square matrices only, although the matrix $\widetilde{M}$ of a temporal network can be non-square. To overcome this issue, we split the original matrix into smaller squares with shorter history and find a linear relationship between the predictability and the number of squares, implying that longer history leads to higher predictability, allowing us to calculate the predictability of any temporal network (see the 'Methods' section and Supplementary Figs 1 and 2 in Supplementary Material, Section III).

## VALIDATION ON MODEL NETWORKS

Next, we test and validate our measure, $\Pi_{\widetilde{M}}^{\max}$, i.e. the topological–temporal predictability (TTP), in synthetic weighted temporal networks (Fig. 2A). The initial snapshot is a network with communities generated by a stochastic block model [26] with links assigned with random weights. In each snapshot henceforth, to generate a neighbor correlation for each link, we activate either the temporal parameter $\gamma$ or the structural parameter $\beta$. With probability $\beta$, we modify the structure and the link changes its weight to that of an adjacent link; with probability $\gamma$, we modify the temporal aspect and the link weight stays the same as in the last snapshot; otherwise, the link is assigned a random value (see the 'Methods' section). Long-range correlations are generated through 2D fractional Gaussian noise (FGN) [27], with a power-law correlation function $C(r,\ \varphi) = r^{-\gamma_x}\cos^2\varphi + r^{-\gamma_y}\sin^2\varphi$, where $(r,\ \varphi)$ are polar coordinates and $\gamma_x$ is regarded as a decay parameter in the temporal dimension, while $\gamma_y$ is for the topological dimension. We compare, in Fig. 2B and C, our TTP with an existing measure, namely temporal predictability (TeP) [28], which considers the links of a temporal network as merely a set of uncorrelated time series and captures only the temporal regularity (see the 'Methods' section), and also with three predictive algorithms. For this, we employ three commonly used methods, namely Markov [29], ConvLSTM [30] and PredNet [31], to forecast the future evolution of real networks. Markov considers a temporal network as a set of uncorrelated time series, ConvLSTM takes into consideration link correlations, and PredNet is a dynamic matrix-prediction algorithm based on ConvLSTM (see Supplementary Material, Section IX for details). As shown in Fig. 2B, TeP is significantly smaller than TTP and, when parameters $\beta$ and $\gamma$ increase, TeP can be seen to be nearly independent of the structural parameter $\beta$, due to the fact that TeP only partially characterizes the regularity of temporal networks. Note that, since $\beta$ and $\gamma$ are not completely independent of each other, TTP is still slightly higher than TeP even when $\beta$ and $\gamma$ approach zero (highest correlations). The higher accuracy of PredNet than the upper bound of predictability provided by TeP, as well as the poor performance of Markov, both indicate the significance of topological information. The unexpected insufficient performance of ConvLSTM, however, is caused by the deconvolution layer, which introduces errors. We further find similar results for 2D FGN, although the varying range of predictabilities is much smaller due to fewer possible values.

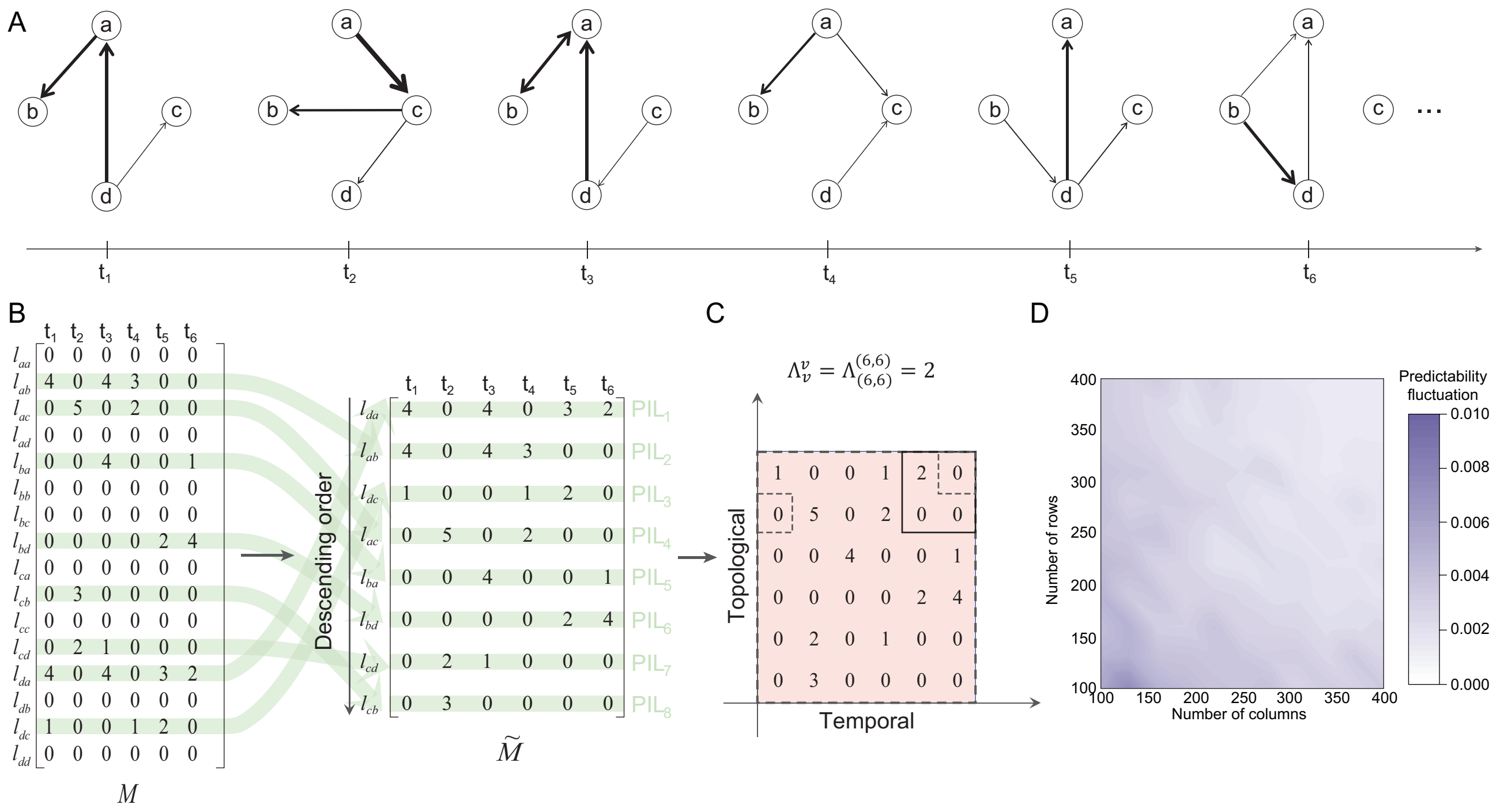

**Figure 1.** Quantifying the predictability of a temporal network. (A) The time-unfolded representation of a temporal network with four nodes. Each snapshot is a weighted directed network where the thickness of links represents their weights. (B) Matrix $M$ encodes the time evolution of each potential link, where each column embodies the structure of a snapshot. Links that rarely appear within the whole duration are removed from the matrix, resulting in matrix $\widetilde{M}$, which captures the meaningful part of $M$ (see the 'Methods' section). The rows of $\widetilde{M}$ are sorted into descending order according to the number of occurrences (see the 'Methods' section). A measure for the predictability of individual links (PIL) that captures only temporal correlations has been developed [28]. (C) Calculation of $\Lambda_v^v$ for a part of $\widetilde{M}$. Note that $\widetilde{M}_{C(k)}$ is defined as a 2D square with side $k$, where $C(k) = \{v = (t,\ s) \in \mathbf{Z}^2 : 0 \leq t \leq k, 0 \leq s \leq k\}$ denotes the coordination set of elements in $\widetilde{M}$, then $\Lambda_v^v \equiv \inf\{k \geq 1 | \widetilde{M}_{u-C(k)} \neq \widetilde{M}_{v-C(k)},\ \forall u \in [\mathbf{0},\ v],\ u \neq v\}$. (D) The fluctuations of topological–temporal predictability (TTP) for different orders of rows in matrix $\widetilde{M}$. All matrices are extracted from a synthetic temporal network in Fig. 2A with rewiring probability $p = 0.5$. We also change $p$ and observe that it has no effect on the results (see Supplementary Material, Section V).

We also introduce synthetic unweighted temporal networks (Fig. 2D) to validate our measure (TTP). The initial snapshot is a ring and each snapshot thereafter is generated by randomly rewiring a fraction $p$ of links in the most recent snapshot. Obviously, as $p$ increases, the network becomes more random, and hence less predictable (see Supplementary Material, Section VI). However, the structural consistency of the aggregated network—a measure that captures only topological regularity on static networks [32]—leads to conflicting increasing predictability (Fig. 2E), demonstrating again the necessity for considering link temporality.

In contrast, our measure, TTP, decreases monotonously when $p$ increases. Yet, due to the high sparsity of these temporal networks, TTP remains high for all values of $p$. To remove the impact of sparsity, we define and calculate the normalized topological–temporal predictability $(\text{NTTP}) = (\text{TTP} - \text{TTP}_{\text{bl}})/(1.0 - \text{TTP}_{\text{bl}})$ for $\text{TTP}_{\text{bl}} < 1.0$ (see the 'Methods' section), where NTTP is the normalized TTP and $\text{TTP}_{\text{bl}}$ is the TTP of the shuffled network, which can be viewed as the lower bound of the predictability of temporal networks. In comparison with NTTP, we also normalize TeP (called here the normalized temporal predictability (NTeP)) over shuffled links (see the 'Methods' section). As shown in Fig. 2F, for $p = 0$, the network is fully predictable $(\text{NTTP} \approx 1.0)$ and, for $p = 1.0$, the network becomes totally random and unpredictable (NTTP vanishes). Even though NTeP has the analogous decreasing behavior, it is usually lower than NTTP due to the lack of topological information. Therefore, the NTTP indeed captures the intrinsic regularity of temporal networks.

## PREDICTABILITY AND PREDICTIVE ALGORITHMS ON REAL NETWORKS

We apply our framework on 18 real temporal networks in diverse scenarios, including animal interactions, human contacts, online communications, political events and transportation (see Supplementary Material, Section I for the description of these network datasets). We group these networks into five categories and reveal the intrinsic predictability profile, consisting of NTTP and NTeP, for each network (Fig. 3A). We find that human contacts have the highest averaged NTTP, probably resulting from their synchronized bursty nature, while temporal regularities dominate the overall predictability of transportation networks due to the periodicity of each link (see Supplementary Material, Section VIII for details). Since the baselines for the normalizations in NTTP and NTeP are different, NTeP can be higher than NTTP. However, we find another interesting phenomenon in most networks (excluding Enron-Email (EE), Levant-Event (LE), Aviation-Network (AN) and Britain-Transportation (BT)): the intrinsic combined predictability is higher than TeP despite the greater complexity of capturing 2D regularity rather than 1D. This implies the significance of the topological information as well as the correlation between the temporal and topological patterns. Surprisingly, we also find strong correlations between topological regularity (characterized by the Hamming distance between each link pair) and the difference between TTP and NPIL (normalized predictability of individual links) (see Fig. 3B and C), suggesting that the intrinsic predictability of real networks mostly originates

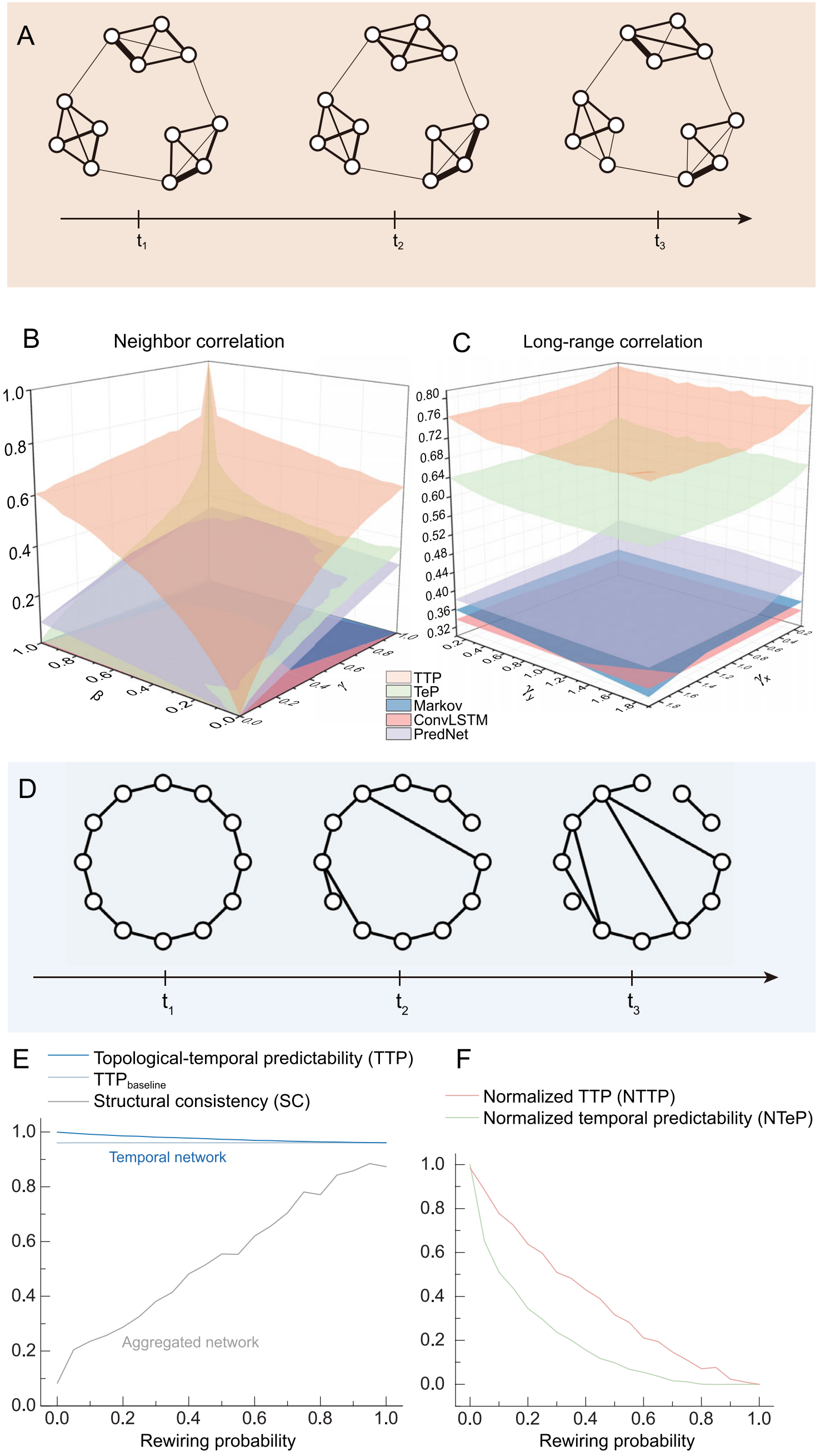

**Figure 2.** Predictability of synthetic temporal networks. (A) To test the impact of link weights on predictability, we develop a temporal stochastic block model with nearest-neighbor correlations (see Supplementary Material, Section VII for details) and long-range correlations [27]. The initial snapshot is generated by the stochastic block model [26], consisting of nodes uniformly assigned to specific communities. There are four communities with 100 nodes and 300 snapshots in each network, while the degree of each node is 3. The link weights in subsequent snapshots are generated according to topological parameters and temporal parameters for the two models, respectively, without changing the network topology. (B) Two predictability measures (TTP and TeP) with three predictive algorithms (Markov [29], ConvLSTM [30] and PredNet [31]) on nearest-neighbor correlations with topological parameter $\beta$ and temporal parameter $\gamma$. Maximum of $\beta$ or $\gamma$ means the strongest memory in the topological or temporal dimension. TeP is obtained by averaging PIL. (C) Two predictability measures with three predictive algorithms on long-range correlations with a power-law correlation function $C(r,\ \varphi) = r^{-\gamma_x}\cos^2\varphi + r^{-\gamma_y}\sin^2\varphi$, where $(r,\ \varphi)$ are polar coordinates and $\gamma_x$ is regarded as the temporal parameter, while $\gamma_y$ is the topological dimension. Results are averaged over 10 independent realizations of the networks. (D) To test the impact of network topology on predictability, we develop an evolving small-world network model. The first snapshot is a ring network; subsequent topologies of the network are generated by randomly rewiring a fraction $p$ of links in the previous snapshot. (E, F) Predictabilities of evolving small-world networks against rewiring probability. The networks are generated by the model in (D) with 50 nodes and average degree 2. Structural consistency (SC) is an existing predictability measure for static undirected and unweighted networks [32]. We normalize TTP over the $TTP_{bl}$ to eliminate the impact of link sparsity (see the 'Methods' section), obtaining the intrinsic predictability of a temporal network, and also obtain normalized TeP for comparison (see the 'Methods' section).

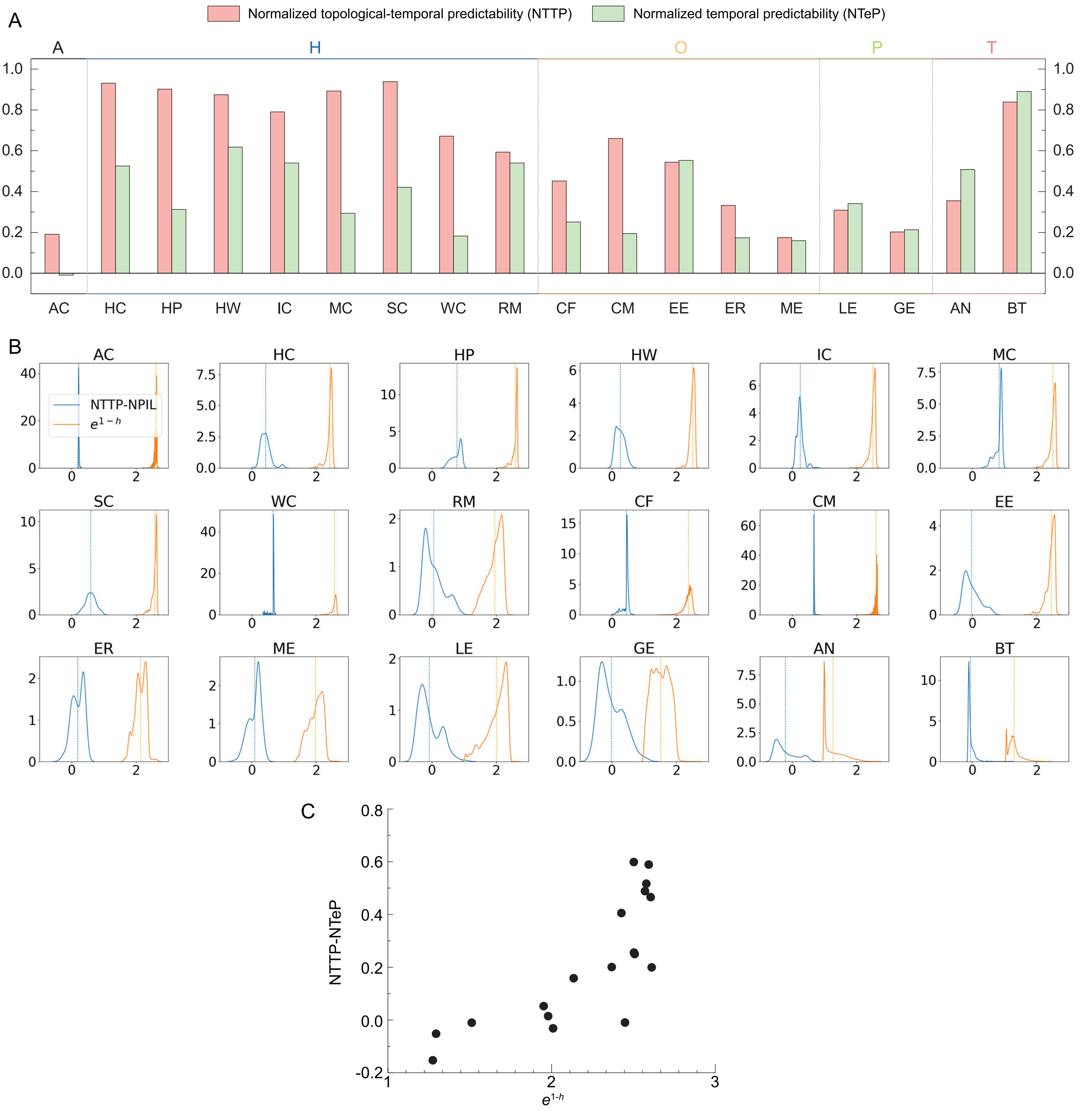

**Figure 3.** Predictability of real temporal networks. (A) NTTP and NTeP for 18 real networks (see Supplementary Material, Section I for the description of these network datasets). 'A' means animal contacts, 'H' denotes human contacts, 'O' means online communications, 'P' represents political events, and 'T' stands for transportation. Note that NTeP can be higher than NTTP because the baselines for these two normalizations are different. (B) Distributions of NTTP-NPIL and $e^{1-h}$ on real-world networks, where NPIL is the normalized predictability of individual links (see the 'Methods' section) and $h$ is the normalized Hamming distance between each link pair. (C) Correlation of average of NTTP-NPIL and $e^{1-h}$ for 18 real networks.

from temporal and topological regularity, rather than from the interdependence between them.

Next, we compare our measure to the predictive power of existing algorithms. We find that the above existing algorithms mostly fall short in prediction (see Fig. 4). Indeed, for a few networks (Ant-Colony (AC): $p$-value $= 7.1 \times 10^{-15}$, College-Message (CM): $p$-value $= 6.1 \times 10^{-7}$), their accuracy is higher than the maximum predictability found by TeP. This is probably because TeP fails to incorporate the topological aspects. However, we found that the predictability given by our TTP measure always remains out of reach from the current algorithms, indicating again that TeP alone cannot characterize the regularities in temporal networks.

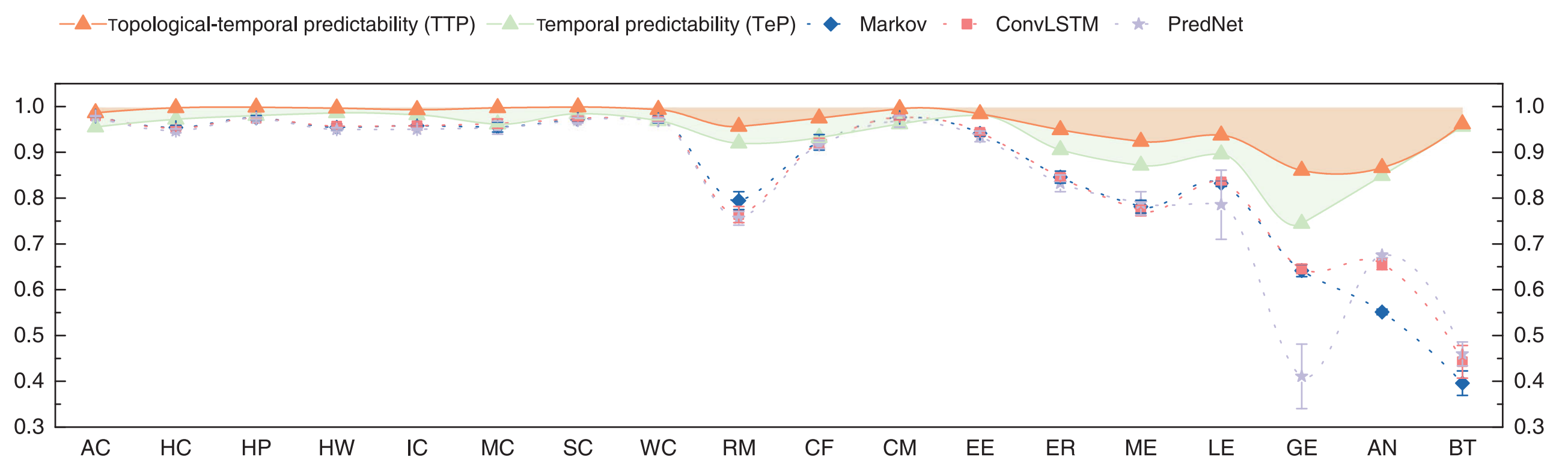

**Figure 4.** Predictive power of existing algorithms. Markov considers a temporal network as a set of uncorrelated time series [29], ConvLSTM takes into consideration of link correlations [30], and PredNet is a dynamic matrix-prediction algorithm based on ConvLSTM [31] (see Supplementary Material, Section IX for details). Error bars are the standard deviation of each algorithm over 10 different runs. Note that all algorithms do not reach our topological–temporal predictabilities of real temporal networks. The accuracy of at least one algorithm is higher than TeP on AC, Marseilles-Contact (MC), Workplace-Contact (WC), College-Forum (CF) and CM networks.

## DISCUSSION AND OUTLOOK

We developed a 2D framework, based on combined topology–temporal features, for quantifying the intrinsic predictability and uncovering the predictability profile of any temporal network. Importantly, we find that the accuracy of current algorithms could be higher than the current temporal-only predictability methods for some real temporal networks. Furthermore, they never exceed our TTP measure. Given the fact that predictability is an essential property of temporal networks, our findings suggest that more accurate predictive algorithms are needed to capture the regularities of real temporal networks, i.e. there is room for researchers to continue improving their predictive algorithms. In addition, applying our measure of predictability to detect the changing points of temporal networks and systematically investigating the impact of predictability on dynamical processes and control on temporal networks are worth future pursuits.

## METHODS

### Matrix filtering

As mentioned above, most rows in matrix $M$ for a real temporal network consistently remain zero. Such link sparsity leads to high predictability. We sort links $\ell_1, \ell_2, \ldots, \ell_n$ in the matrix $M$ by their activation rates $a_1, a_2, \ldots, a_n$ in descending order, hence $a_1 > a_2 > \ldots a_n$, then obtain $\widetilde{M}$ according to the filtering rules:

$$\widetilde{M} = \begin{cases} \{\ell_1, \ell_2, \ldots, \ell_m\} \,|m = \inf\left\{m \in (1, 2, \ldots, n) \,|\, \sum_{i=1}^{m} a_i \geq 0.6 \sum_{i=1}^{n} a_i\right\}, & m < m_\theta \\ \{\ell_1, \ell_2, \ldots, \ell_m\} \,|m = \inf\{m \in (1, 2, \ldots, n) \,|a_m \geq 0.1\}, & m \geq m_\theta \end{cases}.$$

Since the estimation of the entropy rate converges to the real entropy when the size of the matrix goes to infinity [25], we include at least $m_\theta$ most active links or 60% of non-zero elements in the matrix to diminish errors. Due to computational restrictions, we set $m_\theta = 1000$, although, in principle, it could be higher with sufficient resources. All the calculations are performed on $\widetilde{M}$. We show that the matrix-filtering method has no influence on the results after reducing the matrix $M$ to $\widetilde{M}$ in Supplementary Material, Section IV.

### Derivation of predictability and its normalization

Although a detailed derivation is provided in Supplementary Material, Section II, here we adumbrate the main steps used to derive the upper bound of predictability. The entropy rate of a temporal network, which is characterized as a random field, is defined as

$$\begin{aligned} H(M) &\equiv \lim_{\substack{L\to\infty \\ T\to\infty}} \frac{1}{LT} H\left(M^{LT}\right) \\ &= \lim_{\substack{L\to\infty \\ T\to\infty}} \frac{1}{LT} \sum_{\substack{1\le l\le L \\ 1\le t\le T}} H\left(M_{lt}\,|\text{history of } M_{lt}\right) \\ &= \lim_{\substack{L\to\infty \\ T\to\infty}} \frac{1}{LT} \sum_{\substack{1\le l\le L \\ 1\le t\le T}} H(l, t)\,, \end{aligned}$$

where history of $M_{lt} \equiv \{M_{ij} : (j < t)$ or $(j = t$ and $i < l)\}$.

Suppose $P(M_{lt} = \widehat{M}_{lt} \,|\tau_{lt})$ is the probability that is based on the history $\tau_{lt}$, the actual value of $M_{lt}$ agrees with our estimation $\widehat{M}_{lt}$, and $\lambda(\tau_{lt})$ is the probability that $M_{lt}$ takes the most likely value given $\tau_{lt}$, thus

$$\lambda(\tau_{lt}) \equiv \max\left\{P\left(M_{lt} = \widehat{M}_{lt} \,|\tau_{lt}\right)\right\}.$$

Let $P(\tau_{lt})$ be the probability of observing a specific history. It can be demonstrated that the best prediction strategy based on this history is to adopt the most likely value [28]; thus, the predictability of $M_{lt}$ is

$$\Pi_M(l,t) \equiv \sum_{\tau_{lt}} P(\tau_{lt})\lambda(\tau_{lt}).$$

Then the overall predictability $\Pi$ of a random field is

$$\Pi_M \equiv \lim_{\substack{L\to\infty\\ T\to\infty}} \frac{1}{LT} \sum_{\substack{1\le l\le L\\ 1\le t\le T}} \Pi_M(l,t)\,.$$

Because the entropy increases as the distribution becomes uniform, the distribution created by setting the remaining probabilities to be the same while preserving the most likely value $\lambda(\tau_{lt}) = p_{\max}$ has an entropy no less than the original distribution. Note that $M_v \in \mathcal{A}$, $M_v \equiv |\mathcal{A}|$ and denote $N \equiv |\mathcal{A}|$. The entropy of the new distribution is

$$H(M) = -\left(\Pi_M^{\max}\log\Pi_M^{\max} + (1-\Pi_M^{\max})\log(1-\Pi_M^{\max})\right) + (1-\Pi_M^{\max})\log(N-1).$$

Then the solution of $\Pi_M^{\max}$ in the above equation is the upper bound of predictability $\Pi_M$. We adopt the entropy estimator [25] as the entropy rate $H(M)$ for the calculation of predictability's upper bound $\Pi_M^{\max}$

$$H(M) = \frac{n^2\log n^2}{\sum_{v\in C(n)}(\Lambda_v^v)^2}\,,$$

where $v = (v_1, v_2)$, a 2D square with side $k$, is defined as $M_{C(k)}$, $C(k) = \{v = (v_1, v_2) \in \mathbf{Z}^2 : 0 \le v_i \le k, \text{ for all } i\}$ and $\Lambda_v^v$ denotes the smallest integer $k$ such that block $M_{v-C(k)}$ does not occur within the rectangle $(\mathbf{0}, v]$ except at position $v$.

The link sparsity of a temporal network largely determines its predictability even after we adopt matrix filtering. To remove the impact of sparsity and obtain the intrinsic predictability of a temporal network, we normalize $\Pi_M^{\max}$ over the baseline (i.e. predictability of shuffled network), which captures only the regularity in the link-weight distribution. Therefore, we have

$$p_{\text{norm}} = \begin{cases} 1, & b = 1 \text{ and } p = 1 \\ \dfrac{p-b}{(1-b)}, & \text{otherwise} \end{cases},$$

where $p$ is the original predictability and $b$ is the baseline. It is worth noting that the NTeP is the average of the NPIL, which is obtained by normalizing PIL over its own baseline $\text{PIL}_{\text{bl}}$.

## Generalization of predictability using predictive congruency

Since the application of our entropy estimator is limited to only square matrices [25], we explore the correlation of weighted-average predictability and number of squares by gradually splitting the non-square matrix into a set of units, i.e. $1 \times 1$ squares, and compute the predictability of each square. We find the linear relationship between the weighted-average predictability and the number of squares, including units. Assume matrix $\widetilde{M}$ is split into $Q$ squares $s_1, s_2, \ldots, s_Q$ in the first splitting stage, along with $u$ units; let $e_{s_1}, e_{s_2}, \ldots, e_{s_Q}$ be the sizes of squares, then the areas of the squares are $e_{s_1}{}^2, e_{s_2}{}^2, \ldots, e_{s_Q}{}^2$. It is worth noting that we define the predictability of units as $\frac{1}{|\mathcal{A}|}$, where $\mathcal{A}$ is the finite value set of link weights in the temporal network. Then the weighted-average predictability at stage $i$, of which the weight equals the portion of corresponding square in the matrix, is defined as

$$p_i = \left(\sum_{j=1}^{Q-i+1} e_{s_j}{}^2 p_{s_j} + \frac{\sum_{Q-i+2}^{Q} e_{s_j}{}^2 + u}{|\mathcal{A}|}\right)/D,$$

while the number of squares for splitting stage $i$ is

$$N_i = Q - i + 1 + \sum_{j=Q-i+2}^{Q} e_{s_j}{}^2 + u.$$

Note that $N_1 = Q + u$. Since there is a linear relationship between $p_i$ and $N_i$, thus

$$\frac{e_{s_i}{}^2}{e_{s_i}{}^2 - 1}\left(\frac{1}{|\mathcal{A}|} - p_i\right) = kD,$$

$$p_i = kN_i + b,$$

where $1 \le i \le Q-1$, $k$ and $b$ are constants. According to our observations of $k < 0$, the negative linear relationship between $p_i$ and $N_i$ indicates the positive correlation between the length of memory and predictability, since a smaller $N_i$ means larger squares with more memory. We define this as predictive congruency and use it to obtain the TTP of each temporal network with non-square matrix

$$p = p_{N_i=1} = k + b.$$

## Synthetic networks

The temporal stochastic block model is used to test the impact of link weight while topology remains invariant. When generating neighbor correlation in a temporal stochastic block model, we determine the weight of links individually according to parameters $\beta$ and $\gamma$. For each link, there is a probability of modifying the link weight based on structural or temporal aspects. If the structural parameter is selected, then the probability for the link to adopt the same weight as its neighboring link is $\beta$; when the temporal parameter is activated, the probability for the link to remain the same as the previous snapshot equals $\gamma$. Otherwise, the link is assigned a random value. Suppose there are $m$ links $\ell_1, \ell_2, \ldots, \ell_n$ in the matrix; then, the probability density function of a link weight at a certain time is

$$f(\ell_i(t)) = p_\beta \beta \delta_{\ell_i(t)\ell_j(t)} + p_\gamma \gamma \delta_{\ell_i(t)\ell_i(t-1)} + \left(1 - p_\beta \beta - p_\gamma \gamma\right) \delta_{\ell_i(t) r},$$

where $i, j \in [1, n], i \neq j$. $p_\beta$ and $p_\gamma$ are the probabilities to choose the structural parameter $\beta$ and temporal parameter $\gamma$, respectively, $p_\beta + p_\gamma = 1$. $\delta_{xy}$ is the Kronecker delta function, and $r$ is a random number.

Specifically, we assume $p_\beta = p_\gamma = 0.5$ and generate the matrix column by column, from top to bottom within each column. If a link is determined to adopt the weight of its neighboring link at a certain time, we assign the prior link weight to it, which is its adjacent element in the matrix. The initial snapshot of the evolving small-world model is a ring network; then, we obtain each snapshot by rewiring a fraction of links in the previous snapshot.

## TeP and NTeP

TeP is the average PIL in the network (Fig. 1). To eliminate the influence of sparsity, we also normalize PIL over its own baseline to obtain NPIL, and NTeP as the average of NPIL.

$$\text{TeP} = \langle \text{PIL} \rangle,$$

$$\text{NPIL} = \begin{cases} 1, & \text{PIL}_{\text{bl}} = 1 \text{ and PIL} = 1 \\ \dfrac{\text{PIL} - \text{PIL}_{\text{bl}}}{(1 - \text{PIL}_{\text{bl}})}, & \text{otherwise}, \end{cases}$$

$$\text{NTeP} = \langle \text{NPIL} \rangle,$$

where $\text{PIL}_{\text{bl}}$ is the PIL of shuffled links.

## SUPPLEMENTARY DATA

Supplementary data are available at *NSR* online.

## FUNDING

This work was supported by the National Key Research and Development Program of China (2019YFF0301400), the National Natural Science Foundation of China (11875043, 61671031, 61722102 and 61961146005), the Science and Technology Commission of Shanghai Municipality (18ZR1442000) and the Fundamental Research Funds for the Central Universities (22120190251). G.Y. is supported by the National Youth 1000 Talents Program of China.

***Conflict of interest statement.*** None declared.

Supplementary Materials for

# Predictability of real temporal networks

## CONTENTS

# I. Datasets

In the main text we applied our framework to 18 real temporal networks. Ant-Colony (AC) records the interactions between 218 ants (*1*); Aviation-Network (AN) represents the domestic schedule of 224 airports in China; Britain-Transportation (BT) contains the schedules of 5 different types of transportation in the United Kingdom, where the nodes represent the airports or stations and links stand for the traffic flow (*2*); College-Forum network (CF) records user posts and replies in the forum of a university (*3*); College-Message dataset (CM) consists of messages in an online social network at a university (*4*); Enron-Email (EE) collects the emails between 150 employees in Enron Corporation (*5*); European-Research network (ER) is generated using email data from a large European research institution (*6*); Manufacturing-Email network (ME) is an internal email communication network between employees of a mid-sized manufacturing company (*7*); Gulf-Event network (GE) represents political actions between 200 countries and areas; Levant-Event (LE) contains political actions between 462 countries and areas; Haggle-Contact (HC) records the contacts between people measured by carried wireless devices (*8*); Hypertext-Proximity (HP) is a human contact network where each node represents a person and the links between them represent proximity (*9*); Hospital-Ward network (HW) is the temporal network of contacts between patients, patients and workers and among workers in a hospital ward (*10*); Infectious-Contact (IC) records face-to-face behaviors of people during the exhibition INFECTIOUS (*9*); Marseilles-Contact (MC) contains the temporal contacts between students in a high school in Marseilles in 2012 (*11*); Reality-Mining data (RM) is the recording of human contacts among 87 students in MIT (*12*); Student-Contact (SC) is the temporal network of contacts between students in a high school in Marseilles in 2011 (*11*); Workplace-Contact dataset (WC) is the temporal network of contacts between individuals in an office building in France (*13*). The link weight in datasets GE and LE means specific political strategy, while it represents the interaction frequency in other networks.

| Datasets | Category | Type | Nodes | Links | Shape of $\widetilde{M}$ | Link Density | Duration |
|---|---|---|---|---|---|---|---|
| Ant Colony (AC) | A | directed | 278 | 4542 | (1255, 176) | 0.021 | 10S |
| Haggle Contact (HC) | H | undirected | 274 | 11956 | (446, 413) | 0.065 | 10m |
| Hypertext Proximity (HP) | H | undirected | 112 | 3746 | (434, 347) | 0.025 | 10m |
| Hospital Ward (HW) | H | undirected | 75 | 4642 | (183, 573) | 0.044 | 10m |
| Infectious Contact (IC) | H | undirected | 410 | 6612 | (445, 352) | 0.042 | 1m |
| Marseilles Contact (MC) | H | undirected | 180 | 4152 | (453, 200) | 0.046 | 1h |
| Student Contact (SC) | H | undirected | 121 | 3608 | (281, 453) | 0.028 | 10m |
| Workplace Contact (WC) | H | undirected | 92 | 1296 | (160, 273) | 0.03 | 1h |
| Reality Mining (RM) | H | undirected | 96 | 17917 | (392, 227) | 0.201 | 1D |
| College Forum (CF) | O | directed | 899 | 12031 | (1134, 135) | 0.079 | 1D |
| College Message (CM) | O | directed | 640 | 15112 | (4413, 155) | 0.022 | 1D |
| Enron Email (EE) | O | directed | 151 | 11204 | (240, 820) | 0.057 | 1D |
| European Research (ER) | O | directed | 254 | 12748 | (271, 316) | 0.149 | 1D |
| Manufacturing Email (ME) | O | directed | 167 | 33673 | (597, 270) | 0.209 | 1D |
| Levant Event (LE) | P | directed | 462 | 54870 | (1011, 307) | 0.177 | 1M |
| Gulf Event (GE) | P | directed | 200 | 49212 | (532, 244) | 0.379 | 1M |
| Aviation Network (AN) | T | directed | 224 | 840881 | (3407, 365) | 0.676 | 1D |
| Britain Transportation (BT) | T | directed | 147 | 282498 | (1282, 360) | 0.612 | 10m |

**Table S1. Basic properties of the real networks.** In the Category column, ‘A’ means animal interactions, ‘H’ represents human contacts, ‘O’ stands for online communications, ‘P’ represents political events, and ‘T’ means transportation networks. In the duration column, ‘D’ represents day, ‘m’ stands for minute, ‘M’ represents month, ‘S’ means second, while ‘h’ refers to hour.

## II. Predictability

Each temporal network can be considered as a random field on the integer lattice $\boldsymbol{Z}^2$, i.e. a family of random variables $\{M_v : v \in \boldsymbol{Z}^2\}$, indexed by a two-dimensional vector $v = (l, t)$. It is natural to assume that each potential link in the temporal network takes weights in a finite set $\mathcal{A}$, i.e. $M_v \in \mathcal{A}, v \in \boldsymbol{Z}^2$. The uncertainty of a random field can be measured by the entropy rate, which equals the minimum rate required to encode the field without any distortion. Let $M^{LT}$ be the set $\{M_{lt} : 1 \leq l \leq L, 1 \leq t \leq T\}$, the entropy rate of a random field is defined as

$$H(M) \equiv \lim_{\substack{L\to\infty \\ T\to\infty}} \frac{1}{LT} H(M^{LT})$$

$$= \lim_{\substack{L\to\infty \\ T\to\infty}} \frac{1}{LT} \sum_{\substack{1\leq l\leq L \\ 1\leq t\leq T}} H(M_{lt}|history\ of\ M_{lt})$$

$$= \lim_{\substack{L\to\infty \\ T\to\infty}} \frac{1}{LT} \sum_{\substack{1\leq l\leq L \\ 1\leq t\leq T}} H(l,t) \quad \text{(S1)}$$

where $M_{lt}$ is a certain element in matrix $M$. We use $\Omega_{lt}$ to denote the $history\ of\ M_{lt}$, i.e. $\Omega_{lt} = \{M_{ij} : (j < t)\ or\ (j = t\ and\ i < l)\}$, then $H(l,t)$ is equivalent to $H(M_{lt}|\Omega_{lt})$.

To calculate the entropy rate, we use the following estimator which has been demonstrated to have good performance over random fields on the integer lattice $\boldsymbol{Z}^d$ (*14*)

$$\liminf_{n\to\infty} \frac{\sum_{v\in C(n)} (\Lambda_v^v)^d}{n^d \log n^d} \to \frac{1}{H} \quad \text{(S2)}$$

Note that $v = (v_1, v_2, \ldots, v_d)$, a $d$-dimensional cube with side k is defined as $M_{C(k)}$, where $C(k) = \{v \in \boldsymbol{Z}^d : 0 \leq v_i \leq k, \text{for all } i\}$, $\Lambda_v^v$ denotes the smallest integer $k$ such that block $M_{v-C(k)}$ does not occur within the rectangle $(\boldsymbol{0}, v]$ except at position $v$. In our instance, matrix dimension equals two, hence the entropy rate of a temporal network with a large number of snapshots can be estimated as

$$H = \frac{n^2 \log n^2}{\sum_{v \in C(n)} (\Lambda_v^v)^2} \quad \text{(S3)}$$

Let $P(M_{lt} = \widehat{M}_{lt} | \Omega_{lt})$ be the probability that the actual value of $M_{lt}$ agrees with our estimation $\widehat{M}_{lt}$, and $\lambda(\Omega_{lt})$ be the probability that, given $\Omega_{lt}$, $M_{lt}$ takes the most likely value, thus

$$\lambda(\Omega_{lt}) \equiv max\{P(M_{lt} = \widehat{M}_{lt} | \Omega_{lt})\} \quad \text{(S4)}$$

The goal of predictive algorithms is to achieve $\lambda(\Omega_{lt})$, i.e. predicting the most likely value $M_{lt}$ based on $\Omega_{lt}$.

Next, we define predictability $\Pi_M(l,t)$ for a certain element in random field $M$ based on the history. Let $P(\omega_{lt})$ be the probability of observing a specific history $\omega_{lt}$, thus

$$\Pi_M(l,t) \equiv \sum_{\omega_{lt}} P(\omega_{lt}) \lambda(\omega_{lt}) \quad \text{(S5)}$$

The overall predictability $\Pi$ of a random field can be obtained by averaging $\Pi_M(l,t)$ over all elements, i.e.

$$\Pi_M \equiv \lim_{\substack{L \to \infty \\ T \to \infty}} \frac{1}{LT} \sum_{\substack{1 \le l \le L \\ 1 \le t \le T}} \Pi_M(l,t) \quad \text{(S6)}$$

To calculate the upper bound of predictability $\Pi$, inspired by the method in (*15*), we create a new distribution $P(M'_{lt} | \omega_{lt})$ as random as possible for $P(M_{lt} | \omega_{lt})$ while preserving the most likely value $\lambda(\omega_{lt}) = p_{max}$. The rest probabilities are modified to a uniform distribution. Note that $M_v \in \mathcal{A}$, and denote $N \equiv |\mathcal{A}|$. The entropy of the new distribution is

$$H\left(M'_{lt} | \omega_{lt}\right) = -\left(p_{max} log p_{max} + (1 - p_{max}) log(1 - p_{max})\right) + (1 - p_{max}) log(N - 1) \quad \text{(S7)}$$

We assume that $\mathcal{F}(\mathrm{x}) = -\big(x log x + (1 - x) log(1 - x)\big) + (1 - x) log(N - 1)$, thus $H\left(M'_{lt} | \omega_{lt}\right) = \mathcal{F}(\lambda(\omega_{lt}))$ Since $P(M'_{lt} | \omega_{lt})$ is at least as random as $P(M_{lt} | \omega_{lt})$, we have

$$H(M_{lt} | \omega_{lt}) \le H(M'_{lt} | \omega_{lt}) = \mathcal{F}\big(\lambda(\omega_{lt})\big) \quad \text{(S8)}$$

Now we use Equation S1, S5 and the concavity of function $\mathcal{F}(\mathrm{x})$ to obtain the upper bound of predictability $\Pi_M$.

$$H(M) \equiv \lim_{\substack{L\to\infty \\ T\to\infty}} \frac{1}{LT} \sum_{\substack{1\le l\le L \\ 1\le t\le T}} H(M_{lt}|\Omega_{lt}) = \lim_{\substack{L\to\infty \\ T\to\infty}} \frac{1}{LT} \sum_{\substack{1\le l\le L \\ 1\le t\le T}} \sum_{\omega_{lt}} P(\omega_{lt}) H(M_{lt}|\omega_{lt})$$

$$\le \lim_{\substack{L\to\infty \\ T\to\infty}} \frac{1}{LT} \sum_{\substack{1\le l\le L \\ 1\le t\le T}} \sum_{\omega_{lt}} P(\omega_{lt}) \mathcal{F}\big(\lambda(\omega_{lt})\big) \le \lim_{\substack{L\to\infty \\ T\to\infty}} \frac{1}{LT} \sum_{\substack{1\le l\le L \\ 1\le t\le T}} \mathcal{F}\left(\sum_{\omega_{lt}} P(\omega_{lt}) \lambda(\omega_{lt})\right)$$

$$= \lim_{\substack{L\to\infty \\ T\to\infty}} \frac{1}{LT} \sum_{\substack{1\le l\le L \\ 1\le t\le T}} \mathcal{F}\big(\Pi_M(l,t)\big) \le \mathcal{F}\left(\lim_{\substack{L\to\infty \\ T\to\infty}} \frac{1}{LT} \sum_{\substack{1\le l\le L \\ 1\le t\le T}} \Pi_M(l,t)\right) = \mathcal{F}(\Pi_M) \quad \text{(S9)}$$

Since the function $\mathcal{F}(\mathrm{x})$ monotonically deceases with x, we assume $\mathcal{F}(\mathrm{x}) = H(M)$, then the value of x in this equation will be no less than $\Pi_M$, i.e.,

$$H(M) = -\left(\Pi_M^{\max} log \Pi_M^{max} + (1-\Pi_M^{max}) log(1-\Pi_M^{max})\right) + (1-\Pi_M^{max}) log(N-1) \quad \text{(S10)}$$

where the $\Pi_M^{max}$ in the equation is the upper bound of predictability $\Pi_M$, i.e., $\Pi_M^{max} \ge \Pi_M$.

# III. Generalization and predictive congruency

The entropy rate estimator only applies to square matrix, here we generalize the calculation of predictability for general matrices. Specifically, we split a non-square matrix into a set of squares according to the following rules: Start from left or top of the matrix; Split the largest square from the remaining part; Unit cannot be split (see Fig. S1C). This splitting process is executable and unique for any non-square matrix, and the predictability of this non-square matrix is defined as the weighted average predictability of all the square matrices split from it, where the weight of a square equals the proportion of the square area to the original matrix area.

In order to discover the relationship between the number of squares and the predictability, we continue to successively split all squares into units. To ensure the uniqueness of the splitting procedure, we split the closest square matrix to a set of units to obtain the subsequent phase. We calculate again the predictability for current splitting scheme. The process is repeated until the whole matrix is completely split into units.

Suppose the cardinality of finite set $\mathcal{A}$ from which potential links of a temporal network take weights is $|\mathcal{A}|$, the predictability of a unit is then $\frac{1}{|\mathcal{A}|}$, since there is not any historical information anymore.

As shown in Fig. S1A, we find a linear relationship between the predictability $p$ and the number of squares $N$ (Fig. S1A). Since our measure, the topological-temporal predictability (TTP) corresponds to the value of $p$ when there is only one square split from the original matrix, we can obtain TTP by extending the linear relationship to the dot $N$ = 1 (Fig. S1B). We examine the calculation in all other real temporal networks, showing that there are minor deviations in only AN and LE networks (Fig. S6). Suppose matrix $\widetilde{M}$ is split into $Q$ squares $s_1, s_2, \dots, s_Q$, along with $u$ units in the original splitting phase and the square sizes are $e_{s_1}, e_{s_2}, \dots, e_{s_Q}$ respectively, then the predictability acquired at stage $i$ is

$$p_i = \left( \sum_{1}^{Q-i+1} {e_{s_j}}^2 p_{s_j} + \frac{\sum_{Q-i+2}^{Q} {e_{s_j}}^2 + u}{|\mathcal{A}|} \right) / D$$

for $i > 1$, where $D$ is the total area of $\widetilde{M}$. The initial value $p_1 = \left( \sum_1^Q {e_{s_j}}^2 p_{s_j} + \frac{u}{|\mathcal{A}|} \right) / D$, and the number of squares is

$$N_i = Q - i + 1 + \sum_{Q-i+2}^{Q} {e_{s_j}}^2 + u$$

since there is a linear relationship between $p_i$ and $N_i$, we have

$$\frac{p_{i+1} - p_i}{N_{i+1} - N_i} = k$$

where $k$ is a constant and $1 \leq i \leq Q - 1$. The equation leads to

$$\frac{{e_{s_i}}^2}{{e_{s_i}}^2 - 1} \left( \frac{1}{|\mathcal{A}|} - p_i \right) = kD$$

The left side of this equation remains constant for each square in the original splitting phase. We call this property predictive congruency, which is valid for all real and model networks studied in this paper.

We use predictive congruency to obtain the TTP for any non-square matrix. Indeed, using the linear equation $p_i = kN_i + b$, we can estimate TTP by extending it to the dot

$N_i = 1$. According to least square regression

$$k = \frac{(\mathrm{Q}-1)\sum_1^{Q-1}(N_i p_i) - \sum_1^{Q-1} N_i \sum_1^{Q-1} p_i}{(\mathrm{Q}-1)\sum_1^{Q-1} {N_i}^2 - \left(\sum_1^{Q-1} N_i\right)^2}$$

$$b = \frac{\sum_1^{Q-1} p_i - k\sum_1^{Q-1} N_i}{Q-1}$$

hence, TTP for any non-square matrix is

$$p = p_{N_i=1} = k + b$$

$$= \frac{(\mathrm{Q}-1)\sum_1^{Q-1}(N_i p_i) - \sum_1^{Q-1} N_i \sum_1^{Q-1} p_i}{(\mathrm{Q}-1)\sum_1^{Q-1} {N_i}^2 - \left(\sum_1^{Q-1} N_i\right)^2} + \frac{\sum_1^{Q-1} p_i - k\sum_1^{Q-1} N_i}{Q-1}$$

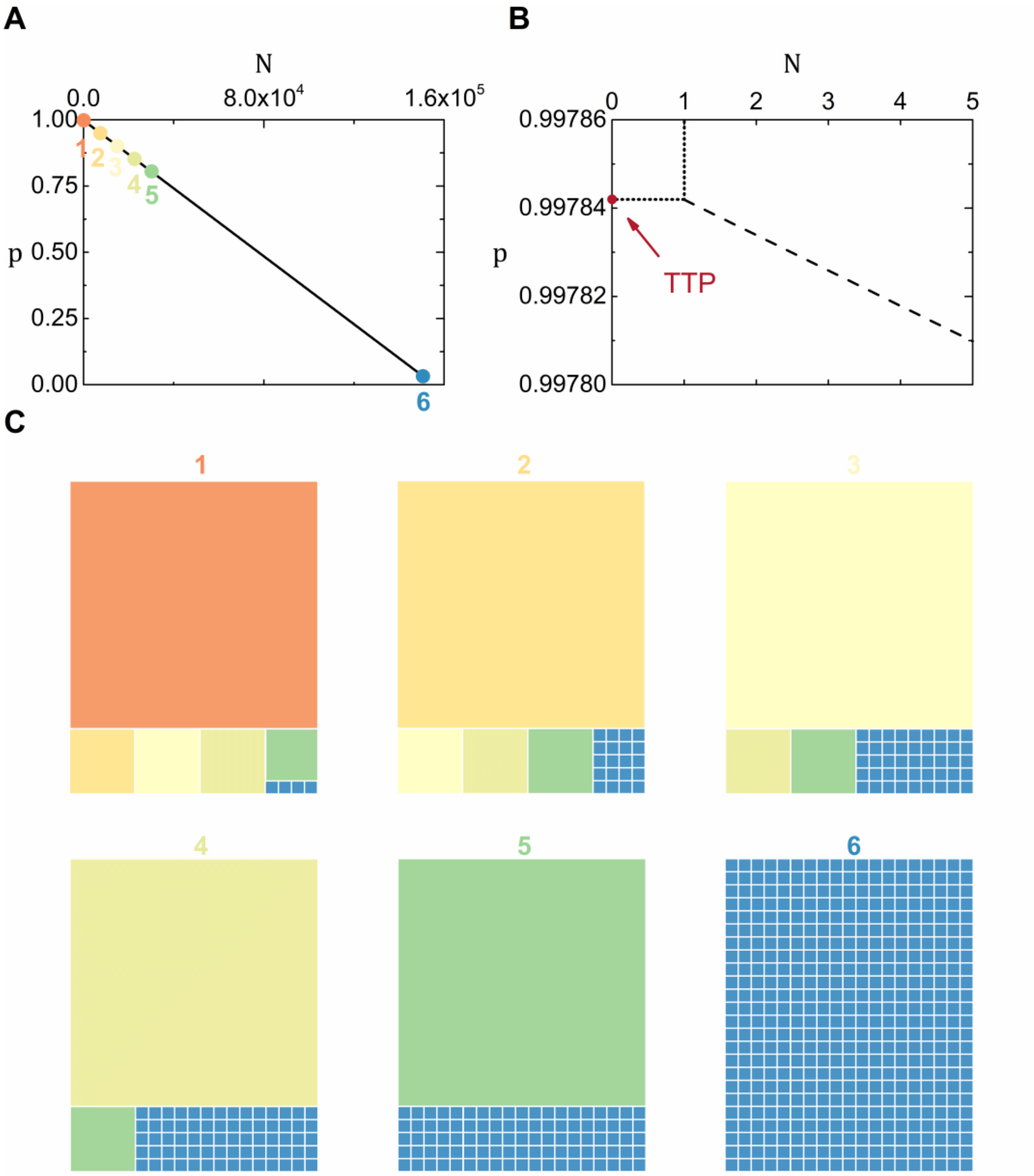

Figure. S1. The splitting process for a non-square matrix and the calculation of predictability. The dataset HP is used as an example. **(A)** N is the number of squares (including units), $p$ is the weighted average predictability of the whole matrix in the splitting phase. Each point corresponds to a splitting phase and the line represents regression. **(B)** The area that approaches axis is zoomed in, where we use the linear relationship to obtain TTP of a matrix. **(C)** Different phases generated by the splitting process. Each matrix corresponds to a splitting phase, the color of each square represents the splitting order: Green denotes the first square to split in next phase, orange is the last square to split while blue represents units. In this instance, we obtain six predictabilities from corresponding splitting phases.

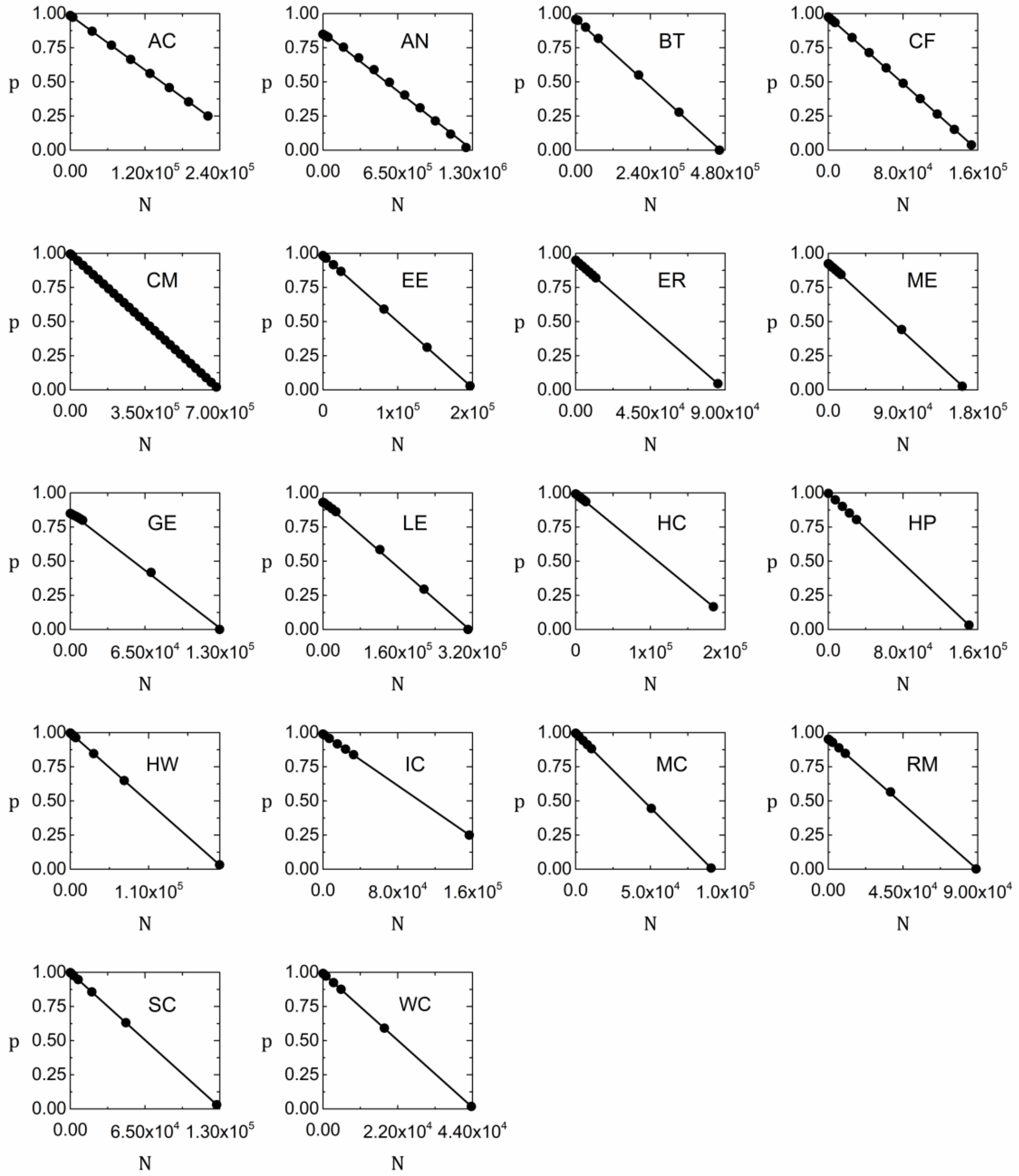

Figure. S2. The relationship between predictability $p$ and the number of squares $N$ for 18 real temporal network datasets. Each point corresponds to a splitting phase and the lines represent respective regression.

# IV. Matrix shuffling and filtering

To obtain $\mathrm{TTP}_{\mathrm{baseline}}$, we usually need to use the average of many different realizations of shuffled network to diminish the errors, yet considering the high time complexity of TTP, we attempt to find a minimum number of realizations that can roughly achieve the same goal. We obtain 100 different shuffling of a network, and average over 100 different combinations of those realizations. Figure S3 indicates the stabilization after we average over 40 different realizations of shuffled network, thus we adopt the average of at least 40 runs as the $\mathrm{TTP}_{\mathrm{baseline}}$ of the network.

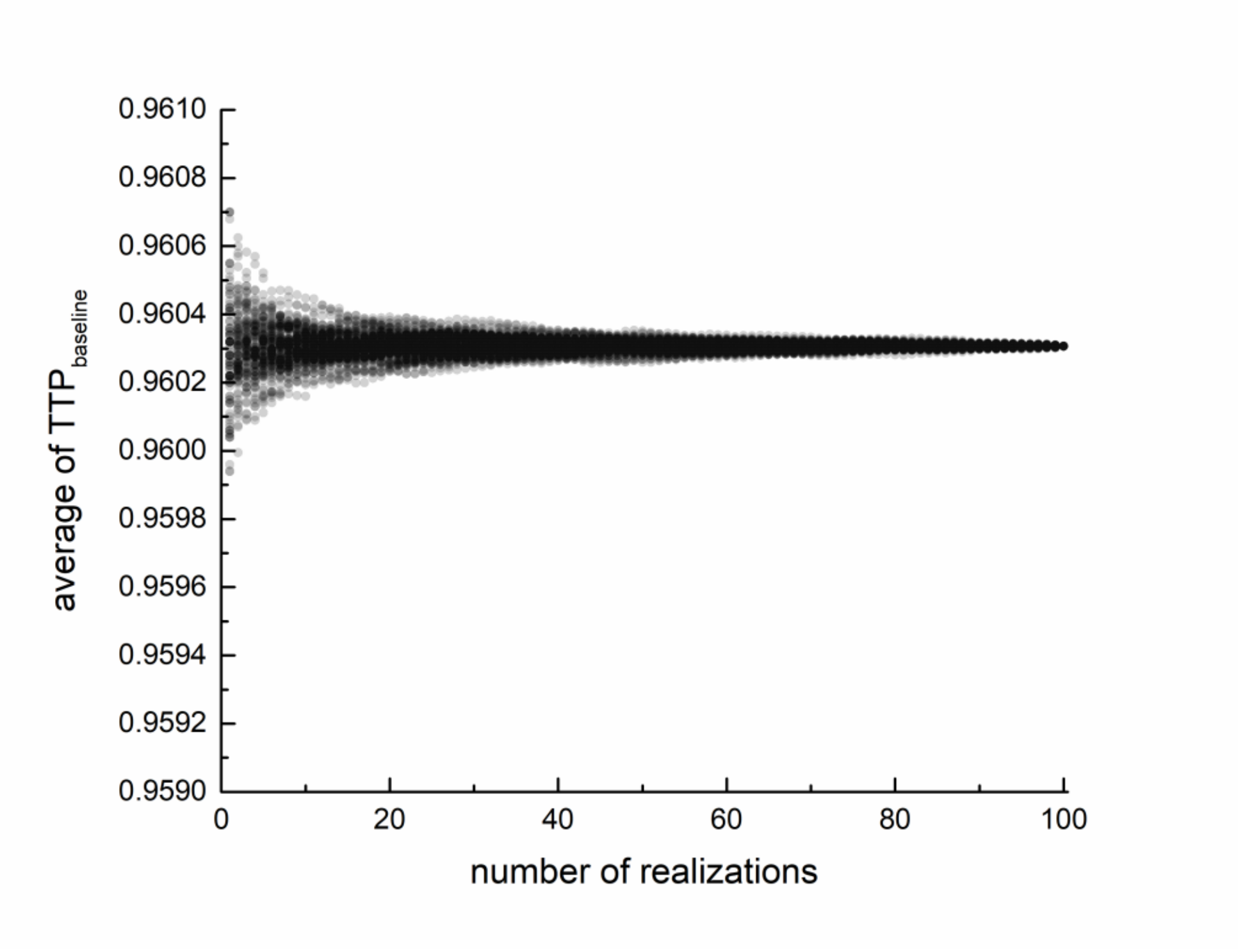

Figure. S3. Average of $\mathrm{TTP}_{\mathrm{baseline}}$ against the number of realizations. A synthetic unweighted network with rewiring probability as 0.1 is used as an example.

Due to the extremely high sparsity of real world networks, we use matrix filtering to remove node pairs that never or seldom have connections (See Methods). By changing the portion of matrix being used, we find that although TTP and $\mathrm{TTP}_{\mathrm{baseline}}$ increase due to more inactive links being included, matrix filtering has no impact on NTTP after 30% of most active links are included in the matrix. We continue to remove inactive links from the remaining matrix of real world networks and find that NTTP maintains stable at an early stage (Sec. VII.).

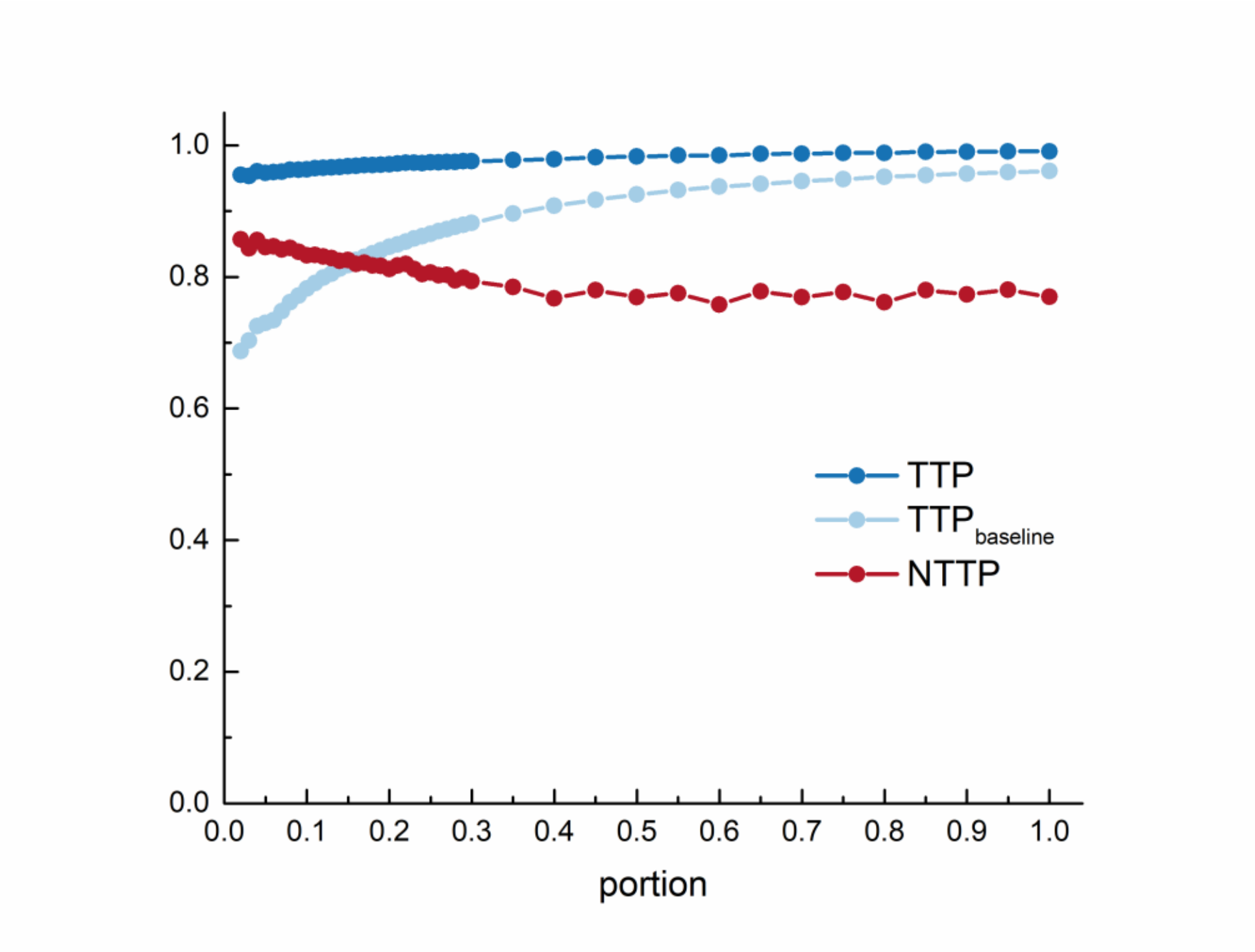

Figure. S4. Impact of portion of active links on TTP, $\mathrm{TTP}_{\mathrm{baseline}}$ and NTTP. A synthetic unweighted network with rewiring probability as 0.1 is used as an example.

# V. Impact of row orders

Different orders of rows in matrix $\widetilde{M}$ might result in different values of entropy rate. In this section, we systematically examine the impact of row orders.

First, we use Genetic Algorithm (GA) to obtain the specific order of rows that achieve the highest or lowest predictability. In GA, the row order is encoded as the chromosome, and the TTP serves as the fitness function (*16*). Assume that there are $n$ rows in the matrix and the ordering sequence is $S = (s_1, s_2, \dots, s_n)$, where $s_i$ represents the sequence number of row $i$ and apparently $s_i \in \{1,2,\dots,n\}; s_i \neq s_j\ for\ i \neq j, i, j \in \{1,2,\dots,n\}$. The genetic sequence is denoted by $G = (g_1, g_2, \dots, g_n)$, and $Y = (y_1, y_2, \dots, y_n)$ represents the ascending sequence $(1, 2, \dots, n)$, then the encoding process is as follows:

1) Suppose $y_j = s_i$, and the sequence number of $y_j$ in $Y$ is $x$, then $g_i = x$;
2) Delete $y_j$ from $Y$;

3) Loop through 1) and 2) until the end.

Since $g_i \in \{1, 2, \ldots, n - i + 1\}$, in the crossover operations we exchange only the corresponding part of two chromosomes, and in mutation operations we mutate only the gene in its value range. Let $D = (d_1, d_2, \ldots, d_n)$ be the decoding sequence, the decoding process is quite similar to encoding:

1) Suppose $y_j$ is the $g_i$-th element in $Y$, then $d_i = y_j$;
2) Delete $y_j$ from $Y$;
3) Loop through 1) and 2) until the end.

The parameters are population size = 50, generation = 100, crossover probability = 0.5, mutation probability = 0.001. In order to retain good individuals with a higher probability in the selection process, we use exponential function to enlarge the difference between good and bad individuals. The goal of genetic algorithm is to obtain not only $\max(p)$, the maximum value of TTP, but also $\min(p)$, the minimum value of TTP. We extract a group of submatrices with different sizes from a real temporal network AN and observe the variations in $\max(p)$ and $\min(p)$, as well as TTP of 100 random orders.

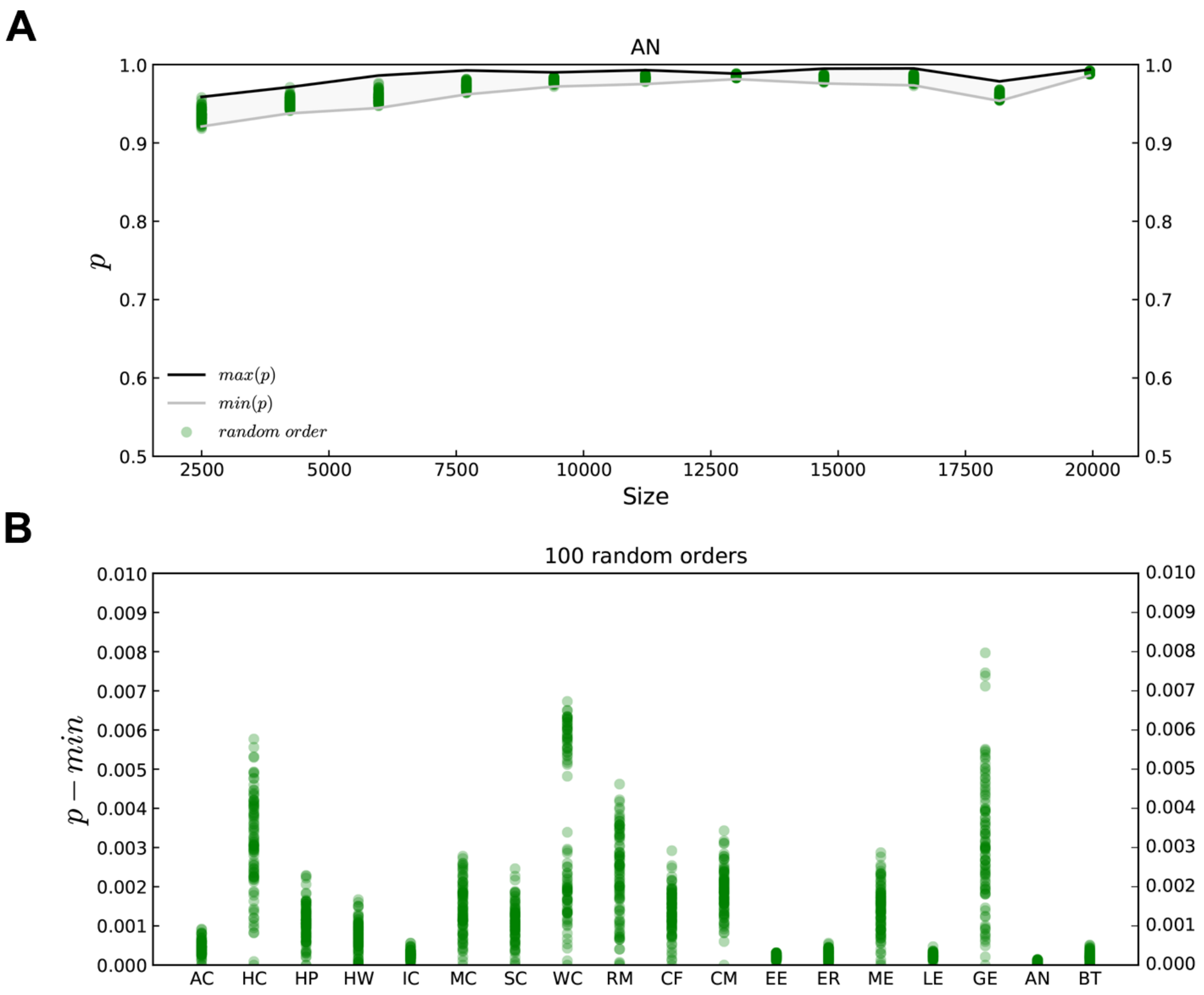

Figure. S5. **(A)** $max(p)$ and $min(p)$ against the size of submatrices extracted from dataset AN. The results of 100 submatrices with random row orders are also provided for comparison. **(B)** We reorder the rows in each dataset of 18 real temporal networks and display the corresponding predictabilities. The results show that the difference in predictability is small (<0.004 for most networks, and <0.008 for all networks).

Figure S5A exhibits the lowest and highest TTPs obtained by genetic algorithm, as well as TTPs for random row orders. We find that the gap between the lowest and the highest TTPs are small and 100 random row orders are enough to sample the TTPs within the gaps. Hence, we sample 100 random row orders for all 18 real temporal networks and test the fluctuations in TTPs. As shown in Fig. S5B, the TTPs of a certain network changes little for different orders of rows in the corresponding matrix. Indeed, for all 18 networks the largest fluctuations are less than 0.008, and for 15 of the networks the largest fluctuations are even less than 0.004. Since TTPs represents the upper bound of predictability of a temporal network, in the paper we calculate TTPs as the maximal predictability in 100 random sorted realizations of the matrix $\widetilde{M}$.

To further test the fluctuation in TTP against row order changes on the first synthetic model, we adjust rewiring probability and observe the similar pattern of variation. Increase in both dimension of size leads to the decline in TTP fluctuation, meanwhile the fluctuation remains consistently lower than 0.008, indicating that TTP is independent of row orders of the matrix.

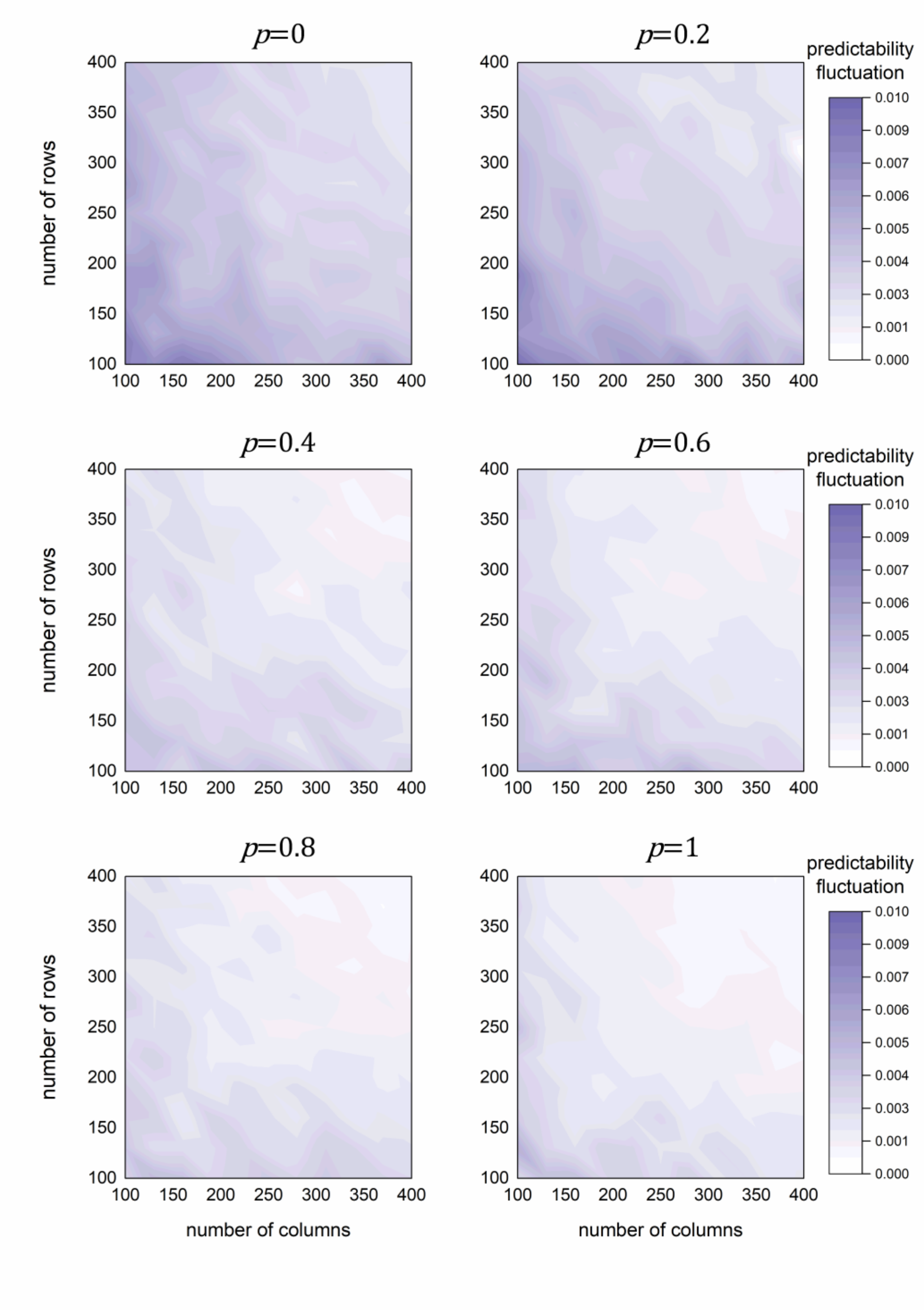

Figure. S6. Fluctuation in TTP on the first synthetic model, with rewiring probability from 0 to 1.0.

# VI. Impact of snapshot duration

To test the impact of time window on TTP and NTTP, we obtain a group of temporal networks by applying different snapshot duration to real datasets.

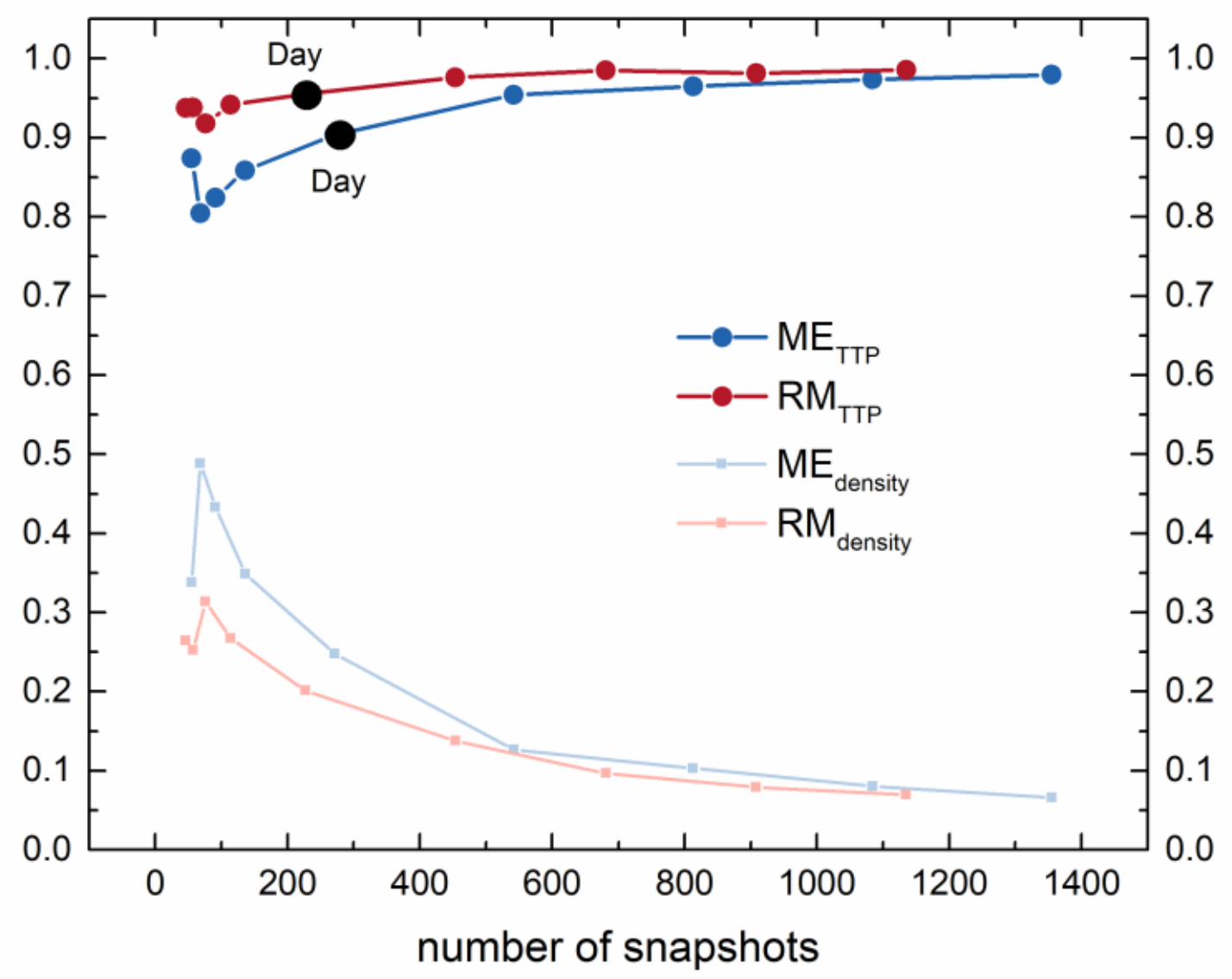

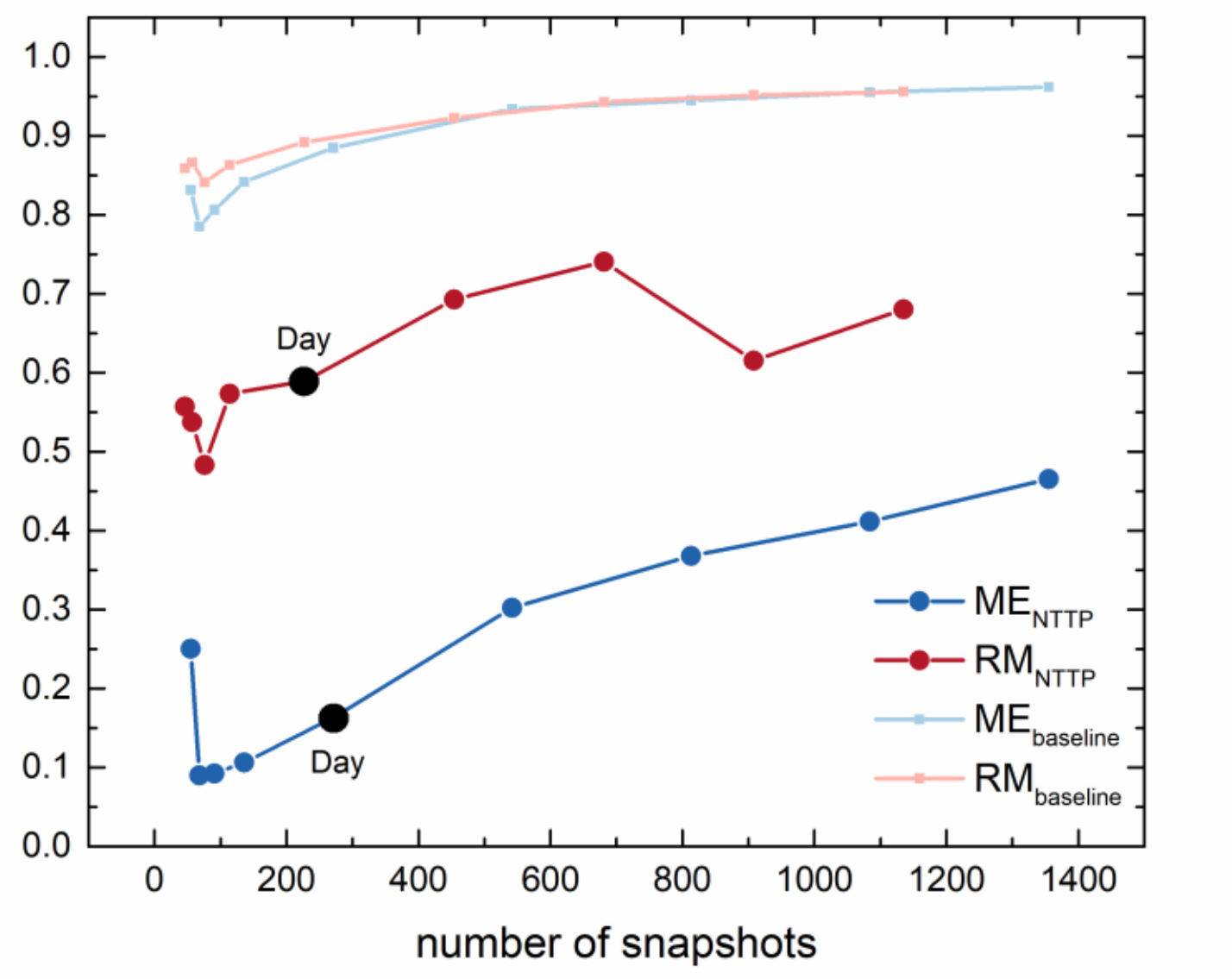

Figure. S7. (Top) TTP and link density vs the number of snapshots. (Bottom) NTTP and baseline vs the number of snapshots.

For datasets ME and RM, there are fluctuations in TTP and NTTP when the snapshot duration changes, but their varying patterns are quite similar (Fig. S7). When snapshot duration becomes smaller and the number of snapshots increases, there is a short decline then a steady increase in TTP. The varying patterns of TTP and NTTP are more similar on ME, while NTTP of RM fluctuates within a small interval. TTP changes in the opposite trend with link density, mainly due to the more predictable nature of more sparse network, especially for the same network with different time windows. Baselines are quite close to TTP in both datasets, implying that network sparsity is the most significant contributor to the high predictability. In this paper, we adopt the most frequently used snapshot duration for each real network.

## VII. NTTP of incomplete data

To test the impact of data incompleteness on the predictability of temporal networks, we remove a proportion of links in each real dataset and calculate its NTTP.

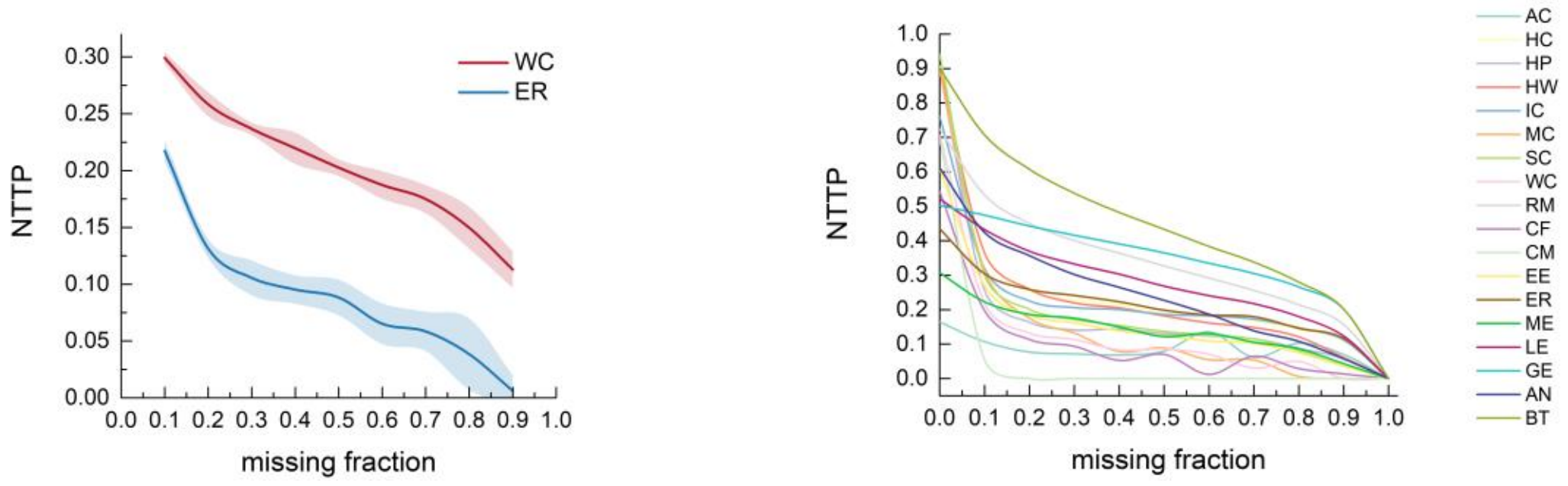

Figure. S8. (Left) NTTP against the fraction of missing links for datasets WC and ER. (Right) NTTP vs missing fraction for all datasets.

As shown in Fig. S8, for most real networks NTTP decreases rapidly with the proportion of links that are removed from datasets. Especially for large networks, such as CM, the predictability decreases to nearly zero when 10% links are removed. The fundamental reason for this phenomenon is that most real temporal networks are extremely sparse. Indeed, each snapshot is usually a sparse network and the expanded matrix that describes all snapshots has only a very small fraction of non-zero entries. The predictability of a temporal network is based only on the patterns of such non-zero entries (existing links). If some of such a small fraction of non-zero entries are removed, the information encoded in the link pattern diminishes quickly. When the network is not too sparse, such as EE, its predictability decreases relatively slowly with the proportion of missing links.

# VIII. NTTP of Submatrices

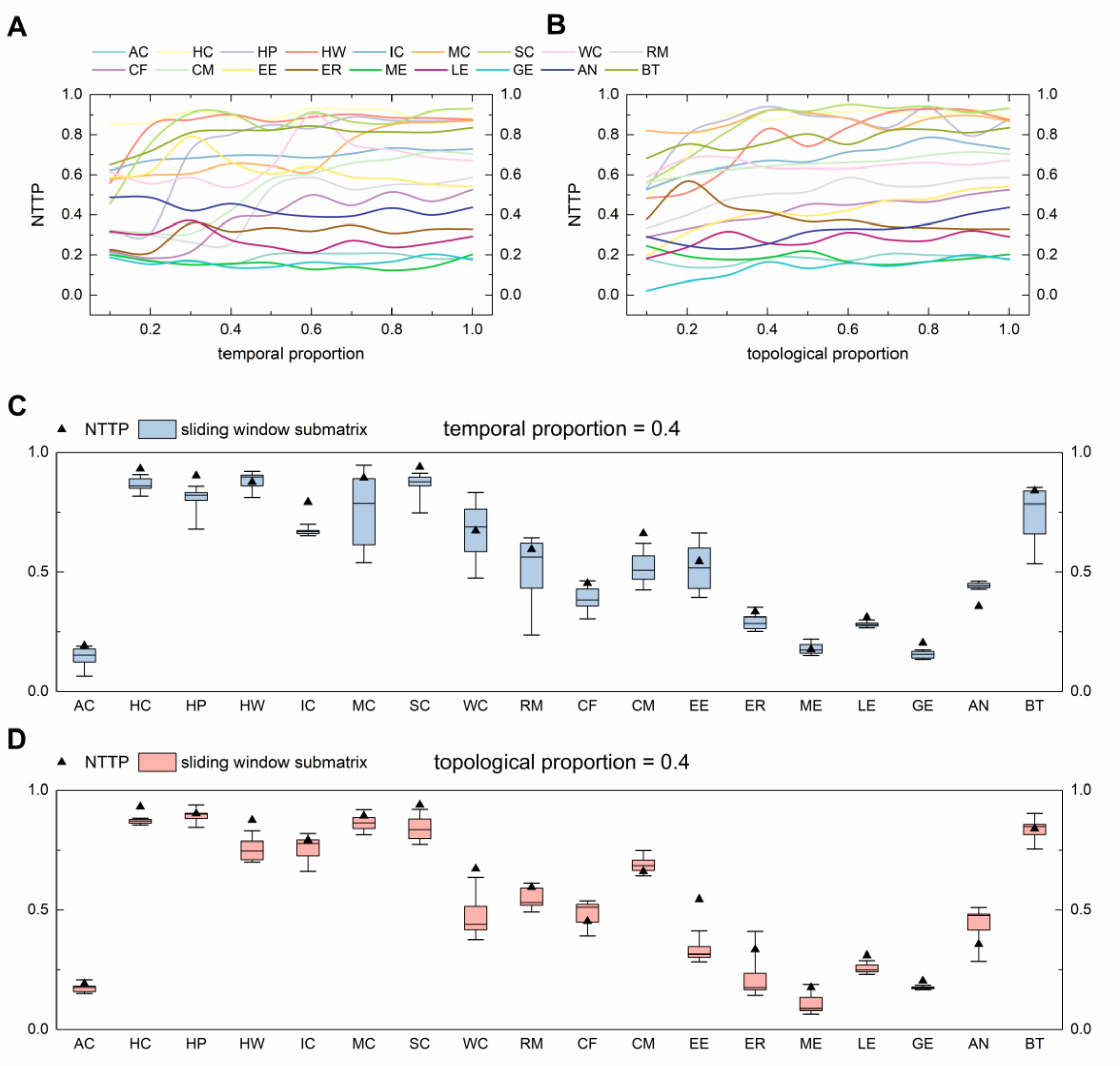

Figure. S9. **(A)** NTTP of a submatrix vs its temporal proportion in the matrix. **(B)** NTTP of a submatrix vs its topological proportion in the matrix. **(C)** NTTP of the whole network and NTTP of sliding window submatrix containing 40% of consecutive observation period. **(D)** NTTP of the whole network and NTTP of sliding window submatrix containing 40% of adjacent links.

In this section we examine the predictability of a part of a temporal network. Figures S9A, B indicate that we can obtain NTTP of the whole network even considering only 70% of adjacent elements in a temporal network or only 50% of the snapshots. Figures S8C, D exhibit the average NTTP while considering temporal submatrices and topological submatrices that contain only 40% snapshots and adjacent links respectively. We find that the NTTPs of 9 (6) real networks are within the predictability intervals of their temporal (topological) submatrices. We further show in Figs. S10B, D that, for networks with large temporal length and link density NTTP can be well estimated through topological submatrices, while for networks with large topological length and link density NTTP can be well estimated through temporal submatrices.

These results indicate that the predictability of a real temporal network can be well estimated from its submatrices. Especially, when the network is extremely large, using submatrices to calculate NTTP can dramatically reduce time complexity at the cost of less than 5% error for all the datasets studied in this paper.

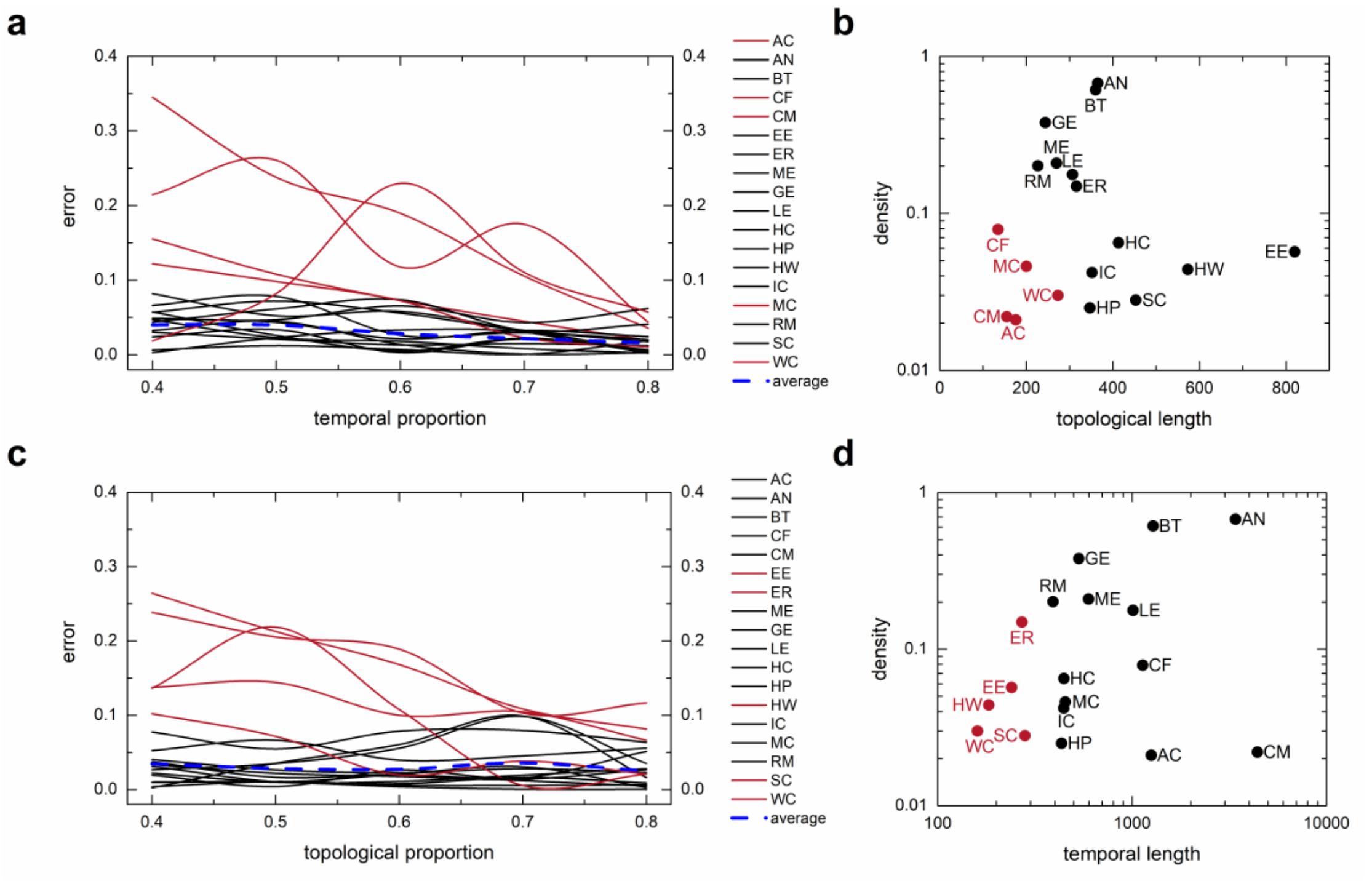

Figure. S10. **(A)** Error of NTTP estimation from temporal submatrix. **(B)** Distribution of datasets in 2-D space indexed by density and topological length. **(C)** Error of NTTP estimation from topological submatrix. **(D)** Distribution of datasets in 2-D space indexed by density and temporal length. Red lines or dots are datasets with large error, black lines or dots refer to networks with relative minor error, and blue lines are mean results of black lines.

# IX. Characteristics of real temporal networks

To determine the cause of high predictability we study the correlation between TTP and $\mathrm{TTP}_{\mathrm{baseline}}$. Interestingly, TTP is proportional to the $\mathrm{TTP}_{\mathrm{baseline}}$ for both model and real networks, indicating that theoretically there is significant room to improve the quantification of predictability. Furthermore, the high predictability mostly comes from the distribution of link weights, i.e. the sparsity of networks, additionally revealing the significance of adopting normalized predictability for analyzing intrinsic predictable nature of temporal networks.

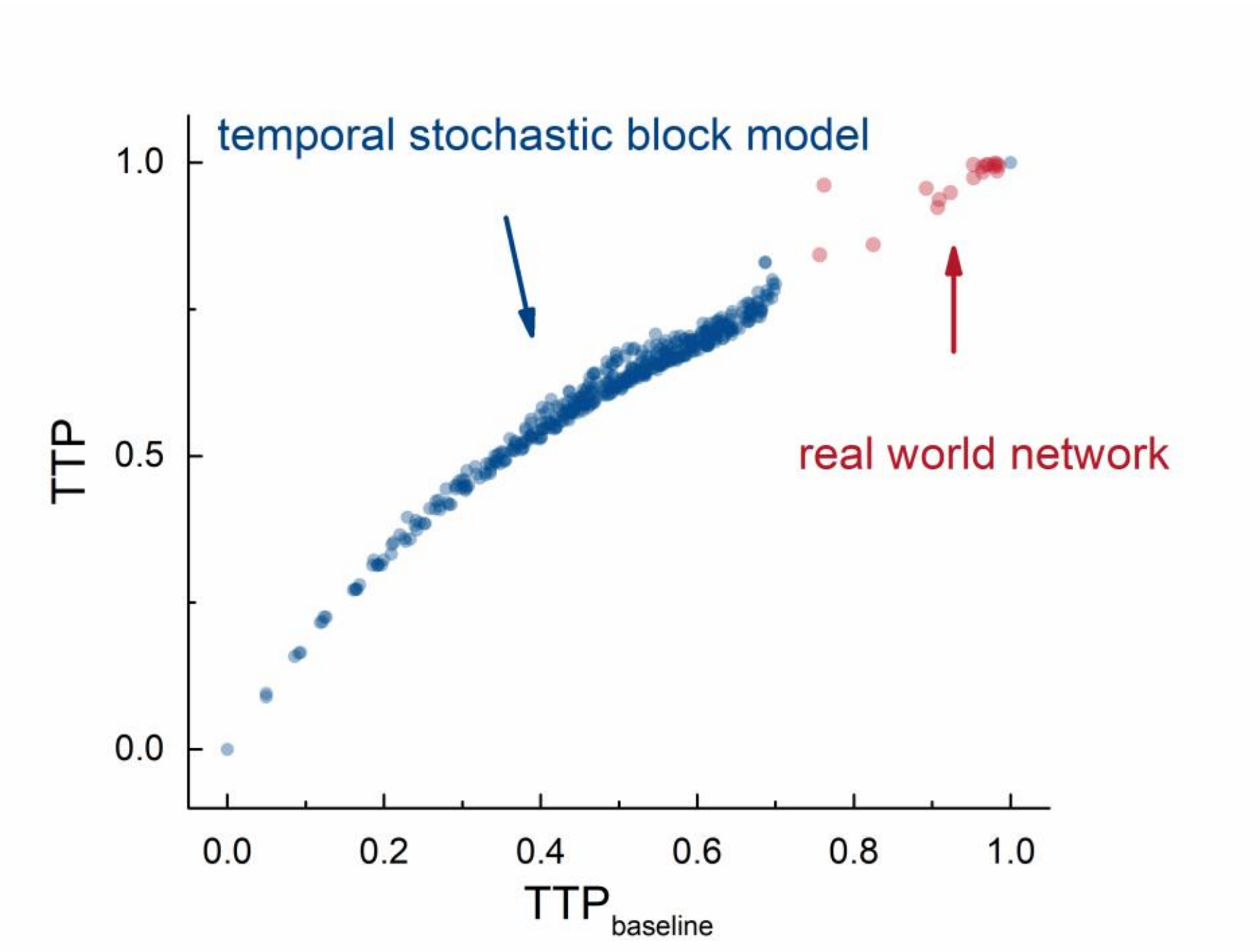

Figure. S11. Correlation between TTP and the $\mathrm{TTP}_{\mathrm{baseline}}$ for real networks as well as temporal stochastic block model.

While there seems to be no significant pattern in animal interactions and political events, activities of human contacts are quite bursty and highly synchronized (Figure S12). It's not difficult to understand the fundamental cause of it lies in the regular nature of human life. Less restricted by space and distance than proximity networks, online communications are considerably less bursty and synchronized, leading to generally lower predictability. Periodicity of BT contributes to its highly predictable nature, while less regular pattern of AN defines its more random essence.

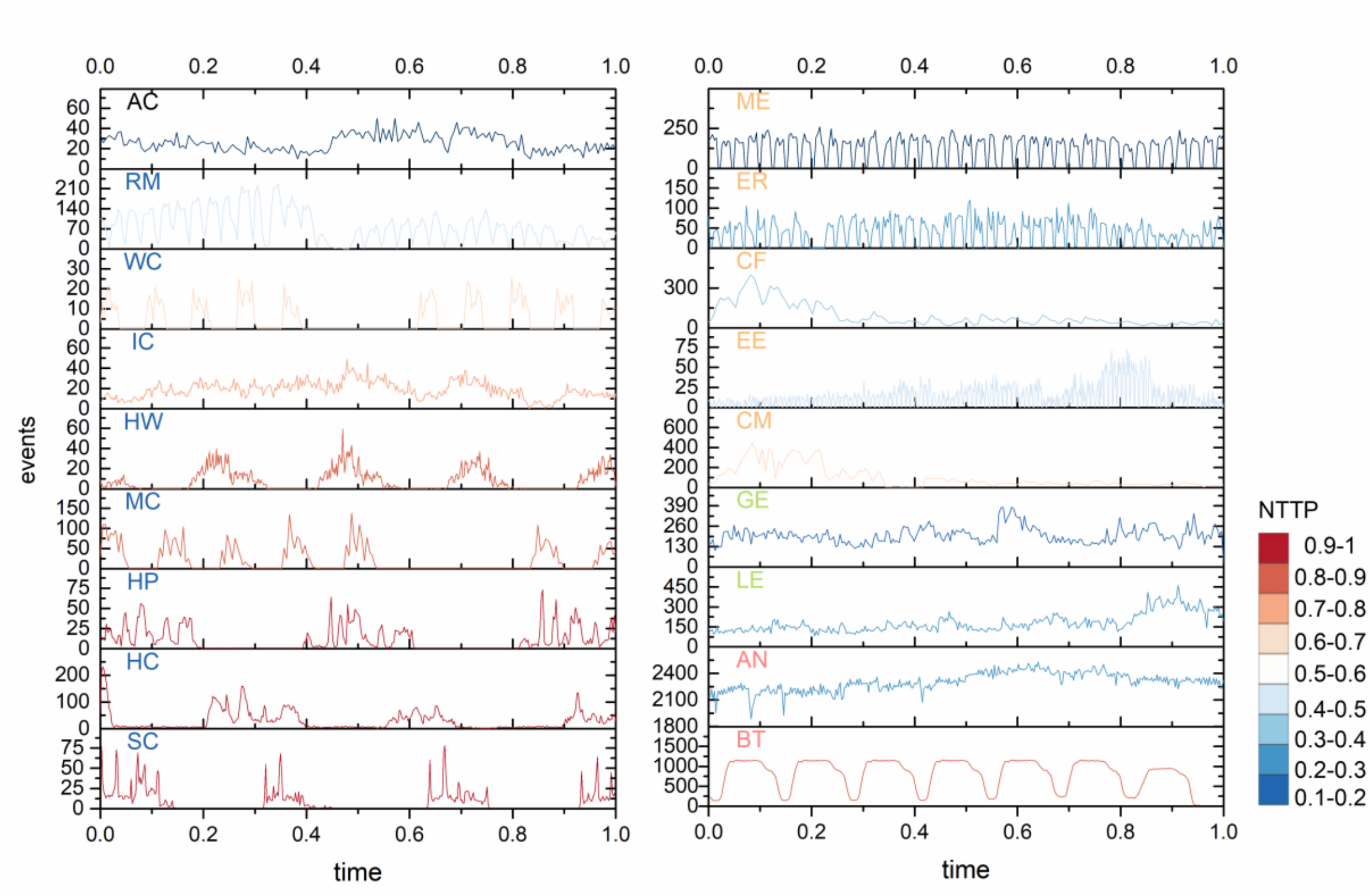

Figure. S12. Number of events in single snapshot with time. Since lengths of time window for networks are different, we use normalized time index for each network. Color of lines corresponds to value of NTTP, while color of legends in each subplot represents its type.

Even though periodic pattern is only observed on human-related networks (human contacts, online communications and transportation) in Figure S12, we discover all the real networks are highly synchronized (See Figure S13), implicating the contacts or events tend to happen at the same time. The symmetry and local maximum also accord perfectly with the observations in Figure S11, strengthening again the periodicity of real networks.

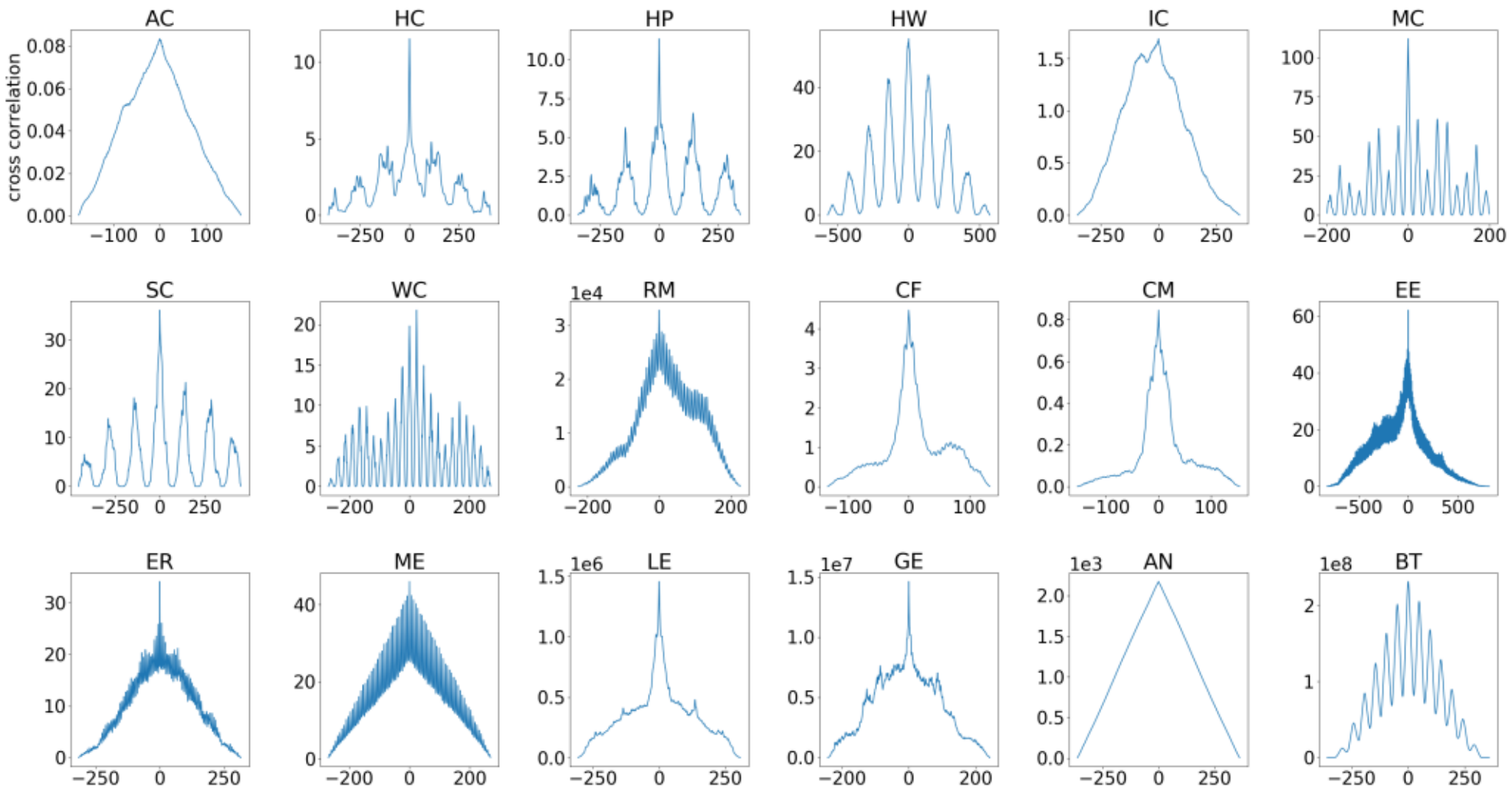

Figure. S13. Cross correlations of all link pairs on real networks. Since the time length of each network can be different, the range of displacement also differs among networks.

Although there's no significant correlation between NTTP and burstiness for real networks (See Figure S14), human contacts have the highest predictability with the most bursty nature among all networks. Based on the characteristics of real networks, burstiness is one of the main causes of the highly predictable nature of human contacts.

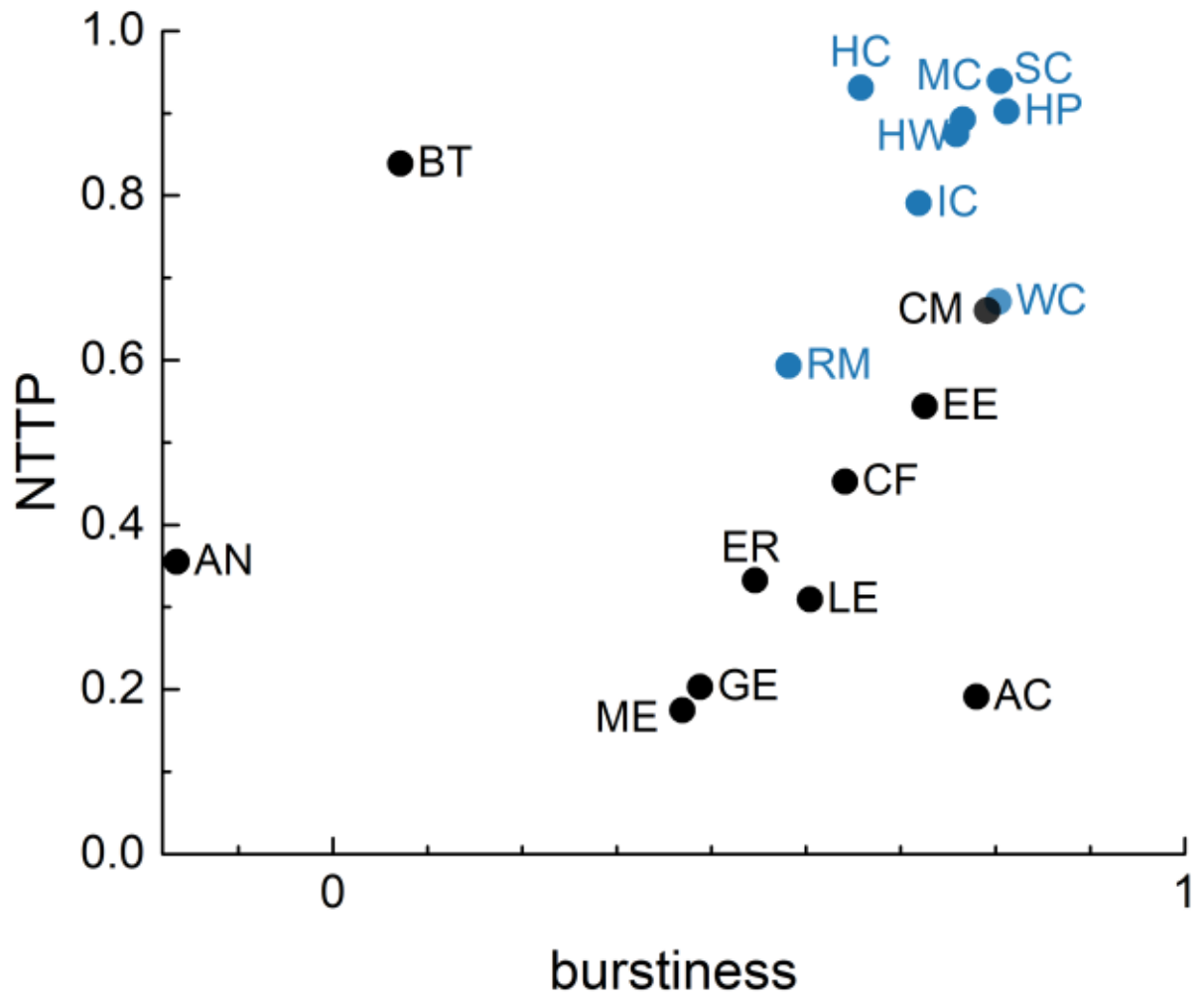

Figure. S14. NTTP vs burstiness on real world networks. Human contact networks are marked as blue dots.

# X. Graphic presentation of model networks

To observe the evolving patterns of model networks introduced in Fig. 2 B and C, we show the visualization of them in Figure S15 and 16. For the neighbor correlation model, the larger the $\beta$ and $\gamma$, the higher the TTP. Since $\beta$ and $\gamma$ are not independently controlling the memory strength, the network becomes much more predictable when they are simultaneously increasing. Despite the fact that $\beta$ and $\gamma$ controls topological and temporal correlation respectively, generated network has displayed diagonal memory. The varying pattern of long-range correlation model is quite similar. The variation trend of TTP corresponds with the regularity of network, demonstrating TTP's validity.

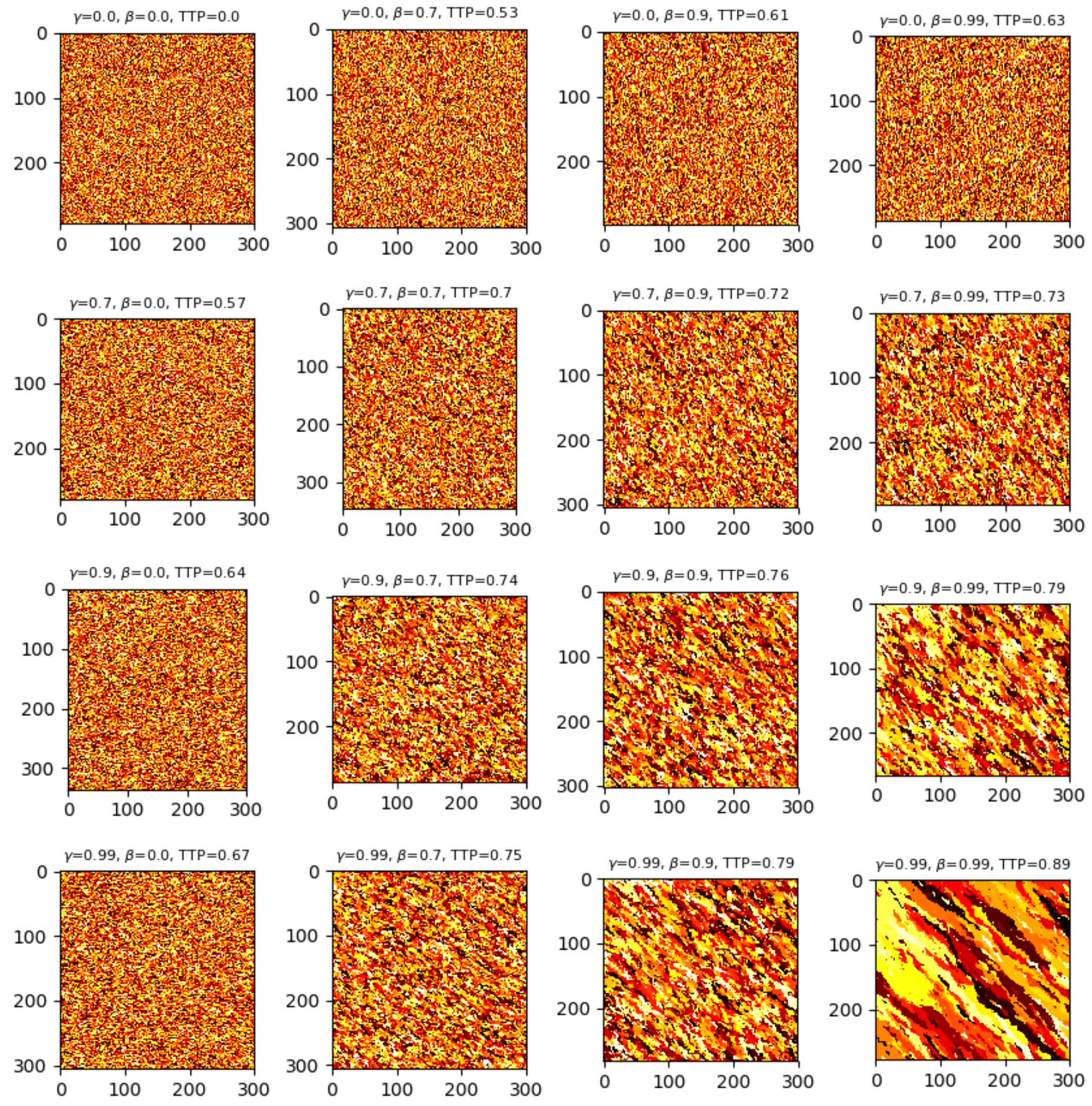

Figure. S15. Visualization of neighbor correlation model. Topological parameter $\beta$ and temporal

parameter $\gamma$ determines the memory strength in each dimension.

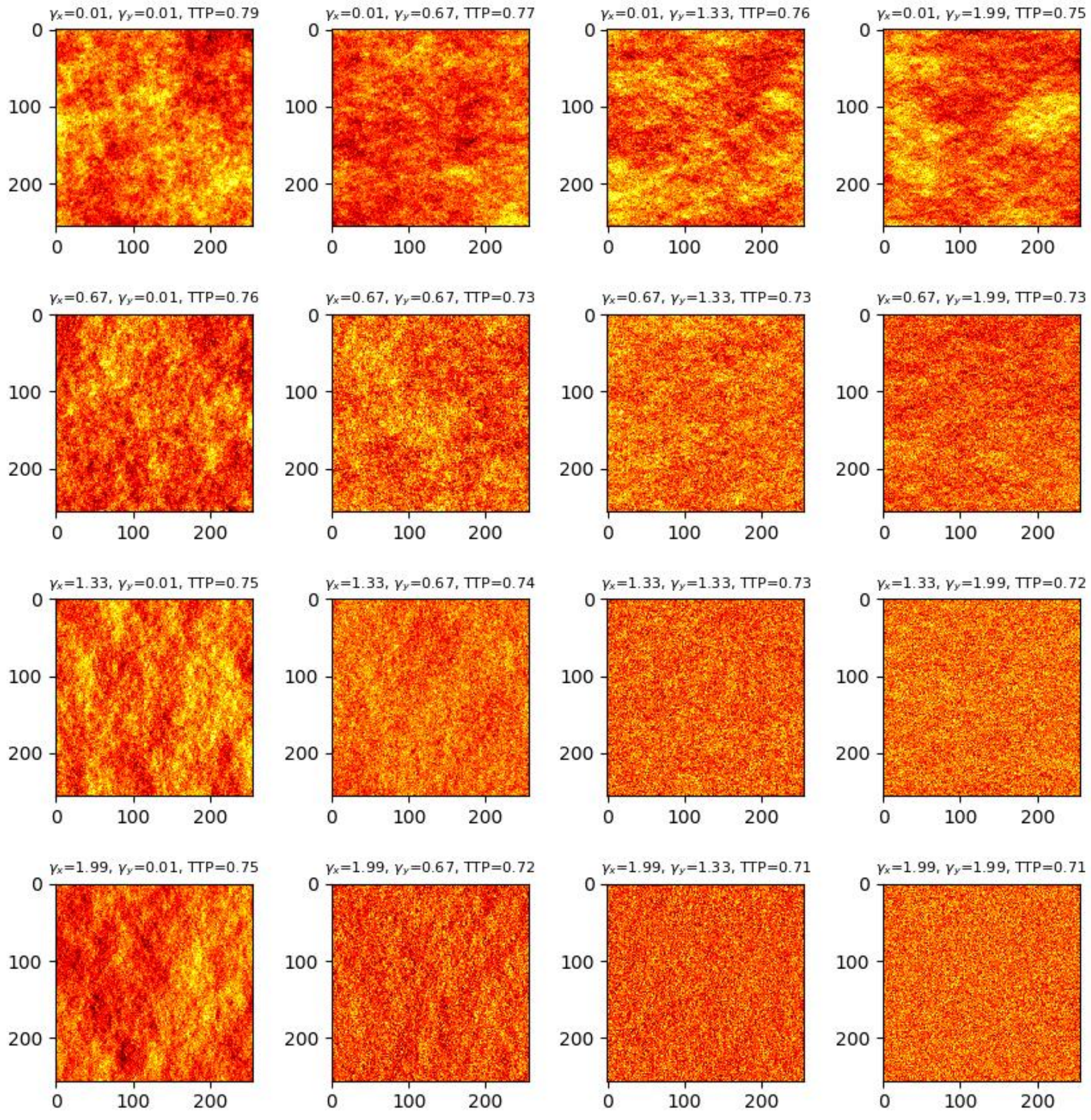

Figure. S16. Visualization of long-range correlation model. $\gamma_x$ and $\gamma_y$ are regarded as decay rates in temporal and topological dimension respectively.

# XI. Predictive algorithms

Markov is a predictive algorithm that considers a temporal network as a set of uncorrelated time series, and only uses the temporal information of each time sequence to predict. Suppose the time series is $S$, its length is $T$, and set the memory length as $l$, then the time series is divided into $T-l$ groups of time series, of which the i-th group consists of $\{S_i, S_{i+1}, \dots, S_{i+l}\}$, with $\{S_i, S_{i+1}, \dots, S_{i+l-1}\}$ being the history to help us predict $S_{i+l}$. These $l$-length groups are randomly shuffled since there are no intergroup correlations under this circumstance, and 70% of the groups are used as training set while 30% are test set.

For each link in the temporal network we construct an $l$-th order transition matrix, in which the rows are the input states of $\{S_i, S_{i+1}, \ldots, S_{i+l-1}\}$, and columns are the output states of the next timestamp $S_{i+l}$. We modify the transition matrix through training set, and use the test set to check its performance. When predicting a time series, previous $l$ symbols prior to the object determine the input state, and the output state with the highest transition probability in the matrix is the output of the algorithm.

The Long Short-Term Memory network (LSTM) (*17*) is considered as an effective model to process sequential data, which is used as a basic model in this paper. The most significant difference between predicting time series and predicting network is that there is some correlation among links in network which could be used to promote prediction. So we adopt convolution LSTM (ConvLSTM) (*18*), an improved version of the LSTM network, which captures spatiotemporal correlations of a network rather than just its temporality.

Assume there are $T$ snapshots in a temporal network, and the history length is set as $s$, which means we use preceding $s$ snapshots to predict the next one. As a result we obtain $T - s$ groups of network packages for prediction. For each network package, former $s$ snapshots are used as the input of ConvLSTM, with the last one as the output, or the label of ConvLSTM, since it is supervised learning. 70% of the network packages is used as the training set and 30% as the test set.

For ConvLSTM, the input $X_1, X_2, \ldots, X_t$, cell outputs $C_1, C_2, \ldots, C_t$, hidden states $H_1, H_2, \ldots, H_t$, and gates $i_t, f_t, o_t$ are three-dimensional tensors, while the counterparts of the FC-LSTM are two-dimensional vectors. Therefore we transform the two-dimensional $1 \times N^2$ matrix into a three-dimensional $1 \times N \times N$ matrix as the input. The ConvLSTM determines the future state of a certain cell in the grid by the inputs and previous states of its local neighbors, which can be easily achieved by using a convolution operator in the state-to-state and input-to-state transitions (see Figure S17).

The ConvLSTM cell is the same as LSTM cell (see Figure S17). The key equations of ConvLSTM are shown below, where '*' denotes the convolution operator and '∘' denotes the Hadamard product:

$$i_t = \sigma(W_{xi} * X_t + W_{hi} * H_{t-1} + W_{ci} \circ C_{t-1} + b_i)$$

$$f_t = \sigma\left(W_{xf} * X_t + W_{hf} * H_{t-1} + W_{cf} \circ C_{t-1} + b_f\right)$$

$$C_t = f_t \circ C_{t-1} + i_t \circ tanh(W_{xc} * X_t + W_{hc} * H_{t-1} + b_c)$$

$$o_t = \sigma(W_{xo} * X_t + W_{ho} * H_{t-1} + W_{co} \circ C_{t-1} + b_o)$$

$$H_t = o_t \circ tanh(C_t)$$

While the key equations of FC-LSTM are shown in below:

$$i_t = \sigma(W_{xi}x_t + W_{hi}h_{t-1} + W_{ci} \circ c_{t-1} + b_i)$$

$$f_t = \sigma\left(W_{xf}x_t + W_{hf}h_{t-1} + W_{cf} \circ c_{t-1} + b_f\right)$$

$$C_t = f_t \circ c_{t-1} + i_t \circ tanh(W_{xc}x_t + W_{hc}h_{t-1} + b_c)$$

$$o_t = \sigma(W_{xo}x_t + W_{ho}h_{t-1} + W_{co} \circ c_{t-1} + b_o)$$

$$h_t = o_t \circ tanh(C_t)$$

We use one hidden layer in ConvLSTM model, and feature preprocess is necessary before training. Feature vectors are normalized using scaler function and transformed through reshape function. We select Relu function as our activation function and Adam algorithm as the optimization function.

Since each link of the network takes value in a discreet set, the output of our predicting algorithm should also be discretized. Therefore before predicting, we discretize our prediction as its closest element in the value set $W$ of the network. Suppose $P$ equals the number of correct predictions, while $N$ is the number of incorrect ones, then the accuracy of prediction for series $m$ is defined as $a_m = \frac{P}{P+N}$, thus the accuracy of predicting the whole network is $a = \frac{\sum_m^M a_m}{M}$.

The Predictive Coding Network (PredNet) (*19*) is a deep convolutional recurrent neural network inspired by the principles of predictive coding from the neuroscience literature. It is trained for next-frame video prediction with the belief that prediction is an effective objective for unsupervised learning. We adopt a three-layer PredNet model as a predictive algorithm in this paper, and the input and data preprocessing are the same as the counterparts of ConvLSTM.

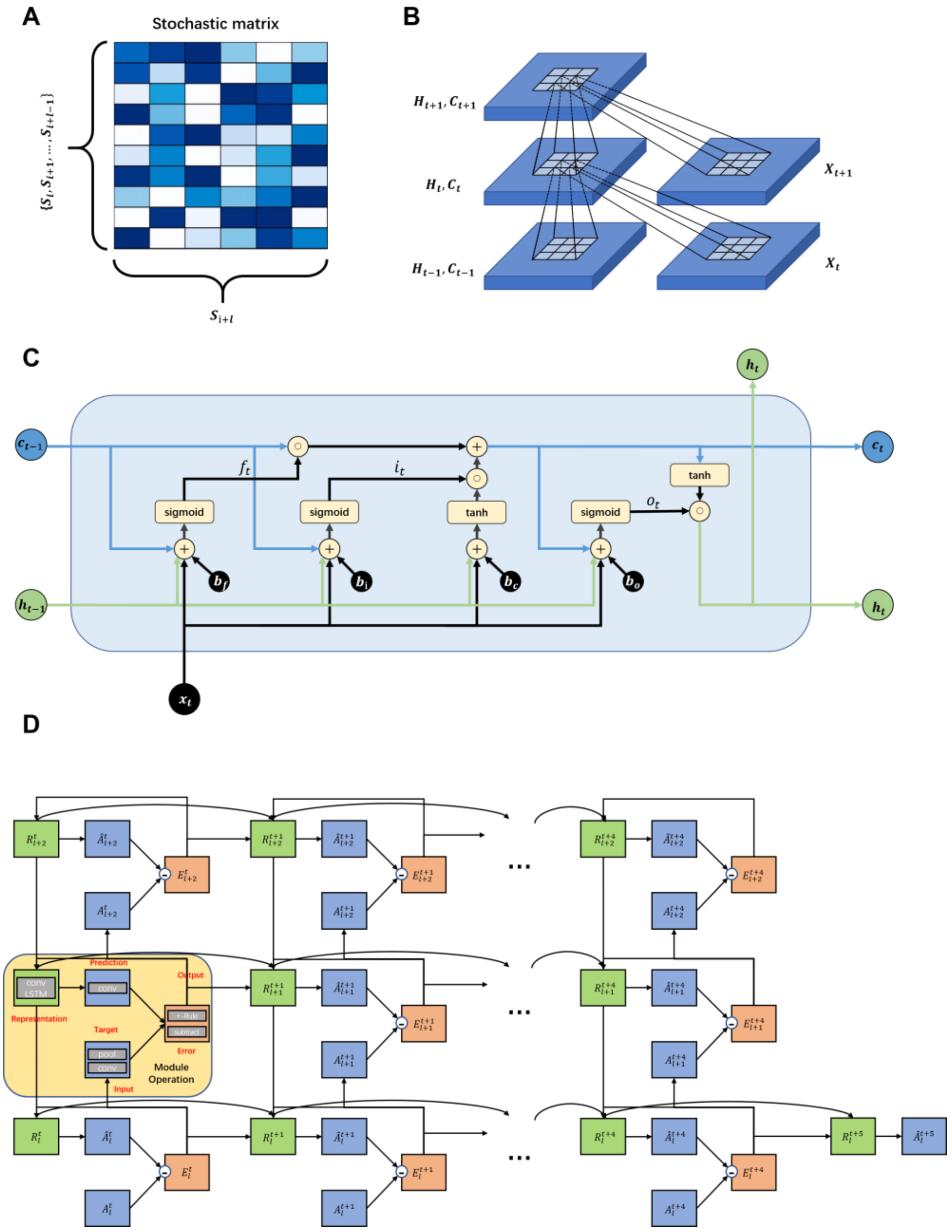

Figure. S17. (a) Transition matrix after training. Each row in the matrix corresponds to an input state vector $\{S_i, S_{i+1}, \ldots, S_{i+l-1}\}$, and each column is the transition probability to an output state $S_{i+l}$. (b) Inner structure of ConvLSTM, $X$ is the input while $H$ and $C$ are the parameters. (c) Structure of ConvLSTM cell. (d) Flow diagram of PredNet. $S_i, S_{i+1}, \ldots, S_{i+l-1}$ are the input states and $S_{i+l}$ is the predicted state.